\documentclass[usenatbib]{mn2e}
\usepackage{lscape}
\usepackage{amsmath,amssymb}
\usepackage{mathrsfs}	
\usepackage[dvipdfmx]{graphicx}

\usepackage{epstopdf}

\usepackage{txfonts}
\usepackage[T1]{fontenc}
\usepackage{aecompl}

\catcode`\@=11
\def\gta{\ifmmode{\,\mathrel{\mathpalette\@versim>\,}}
    \else{$\,\mathrel{\mathpalette\@versim>}\,$}\fi}
\def\lta{\ifmmode{\,\mathrel{\mathpalette\@versim<\,}}
    \else{$\,\mathrel{\mathpalette\@versim<}\,$}\fi}
\def\@versim#1#2{\lower 2.9truept \vbox{\baselineskip 0pt \lineskip
    0.5truept \ialign{$\m@th#1\hfil##\hfil$\crcr#2\crcr\sim\crcr}}}
\catcode`\@=12  

\makeatletter
\def\blfootnote{\xdef\@thefnmark{}\@footnotetext}
\makeatother
\newcommand{\ds}{\displaystyle}

\newcommand{\de}{{\rm d}}

\newcommand{\Jr}{J_{\rm r}}

\newcommand{\bx}{{\bf x}}

\newcommand{\sigmar}{\sigma_{\rm r}}

\newcommand{\vr}{v_{\rm R}}

\newcommand{\Vc}{v_{\rm c}}

\newcommand{\mE}{\mathcal{E}}
\newcommand{\mP}{\mathcal{P}}

\newcommand{\vy}{{\bf y}}
\newcommand{\rb}{r_{\rm b}}
\newcommand{\rh}{r_{\rm h}}

\renewcommand{\[}{\begin{equation}}
\renewcommand{\]}{\end{equation}}

\def\bmu{{\bi\mu}}\def\btheta{{\bi\theta}}

\let\boldgrk=\gkvecten
\let\boldgrksc=\gkvecseven

\def\gkthing#1{{\mathchoice%
	{\hbox{{\boldgrk\char#1}}}
	{\hbox{{\boldgrk\char#1}}}
	{\hbox{{\boldgrksc\char#1}}}
	{\hbox{{\boldgrksc\char#1}}}}}

\def\vtheta{\gkthing{18}}

{\newif\ifnotend
\notendtrue
\def\veclist{ABCDEFGHIJKLMNOPQRSTUVWXYZabcdefghijklmnopqrstuvwxyz.}
\def\top#1#2.{#1}
\def\tail#1#2.{#2.}
\loop\expandafter\xdef\csname v\expandafter\top\veclist\endcsname%
{{\noexpand\bf\expandafter\top\veclist}}
\edef\veclist{\expandafter\tail\veclist}
\if\veclist.\notendfalse\fi\ifnotend\repeat}
\def\df{{\sc DF}}

\def\cE{{\cal E}}
\def\fracj#1#2{{\textstyle{#1\over#2}}}
\def\pa{\partial}

\def\e{{\rm e}}
\def\d{{\rm d}}

\defcitealias{Binney2014}{B14}

\begin{document}

\date{Accepted 2014 December 5.  Received 2014 December 5; in original form 2014 November 28}
\title[Action-based models for spheroidals]
{Action-based distribution functions for spheroidal galaxy components}
\author[L. Posti et al.]{Lorenzo Posti$^{1}$\thanks{E-mail: lorenzo.posti@unibo.it},
James Binney$^{2}$, Carlo Nipoti$^{1}$ and Luca Ciotti$^{1}$
\\ \\
$^{1}$Dipartimento di Fisica e Astronomia, Universit\`a di Bologna, viale Berti-Pichat 6/2, I-40127 Bologna, Italy \\
$^{2}$Rudolf Peierls Centre for Theoretical Physics, Keble Road, Oxford OX1 3NP, UK}

\maketitle

\begin{abstract}
We present an approach to the design of distribution functions that depend on
the phase-space coordinates through  the action
integrals. The approach makes it easy to construct a dynamical model of a given
stellar component. We illustrate the approach by deriving distribution
functions that self-consistently generate several popular stellar systems,
including the Hernquist, Jaffe, and Navarro, Frenk and White models.
We focus on non-rotating spherical systems, but extension to flattened and
rotating systems is trivial. Our distribution functions are
easily added to each other and to previously published distribution functions for discs to
create self-consistent multi-component galaxies. The models this approach makes
possible should prove valuable both
for the interpretation of observational data and for exploring the
non-equilibrium dynamics of galaxies via N-body simulations. 
\end{abstract}

\section{Introduction}

Axisymmetric equilibrium models are extremely useful tools for the study of
galaxies. A real galaxy will never be in perfect dynamical equilibrium -- it
might be accreting dwarf satellites, or being tidally disturbed by the
gravitational field of the group or cluster to which it belongs, or
displaying spiral structure -- but an axisymmetric equilibrium model will
usually provide a useful basis from which a more realistic model can be
constructed by perturbation theory.

By \cite{Jeans1915} theorem, every equilibrium model can be described by a
distribution function (\df) that depends on the phase-space coordinates
$(\vx,\vv)$ only through isolating integrals of motion. In an axisymmetric
potential, most orbits prove to be quasiperiodic, with the consequence that
they admit three isolating integrals \citep{Arnold}. Consequently, a generic \df\ for an
axisymmetric equilibrium galaxy is a function of three variables.

The major obstacle to exploiting this insight is that we have analytic
expressions for only two isolating integrals of motion in a general
axisymmetric potential, namely the energy $E=\fracj12v^2+\Phi(\vx)$ and the
component of the angular momentum about the symmetry axis,
$J_\phi=(\vx\times\vv)_z$. Several authors have examined model galaxies with
\df s of the two-integral form $f(E,J_\phi)$
\citep{PrendergastTomer,Wilson75,Rowley88,Evans1994}, but in such models the
velocity dispersions $\sigma_R$ and $\sigma_z$ in the radial and vertical
directions are inevitably equal. This condition is seriously violated in
our Galaxy and we have no reason to suppose that the condition is better
satisfied in any external galaxy. Hence it is mandatory to extend the \df's
argument list to include a ``non-classical'' integral, $I_3$, for which we do
not have a convenient expression.

Since any function $J(E,J_\phi,I_3)$ of three isolating integrals is itself
an isolating integral, we actually have an enormous amount of freedom as to
what integrals to use as arguments of the \df. Given that we must use at
least one integral for which we lack an expression for its dependence on
$(\vx,\vv)$, there is a powerful case for making the \df's arguments
action integrals. These integrals are alone capable
as serving as the three momenta $J_i$ of a canonical coordinate system -- this
property makes them the bedrock of perturbation theory. Their canonically
conjugate variables, the angles $\theta_i$, have two remarkable properties:
(i) along any orbit they increase linearly with time at rates $\Omega_i(\vJ)$, so
 \[
\theta_i(t)=\theta_i(0)+\Omega_i(\vJ)\,t,
\]
 and (ii) they make the ordinary phase-space coordinates periodic functions
\[
\vx(\vtheta+2\pi\vm,\vJ)=\vx(\vtheta,\vJ)\quad (\hbox{integer }m_i).
\]
The actions $J_i$ also have nice properties. In particular, (i) any triple of
finite numbers $(J_r,J_\phi,J_z)$ with $J_r,J_z\ge0$ corresponds to a bound
orbit with the orbit $\vJ=0$  being that on which a star is stationary at the
middle of the galaxy, and (ii) the volume of phase space occupied by orbits
with actions in $\d^3\vJ$ is $(2\pi)^3\d^3\vJ$. Consequently,
any
non-negative function $f(\vJ)$ that tends to zero as $|\vJ|\to\infty$ and has
a finite integral $\int\d^3\vJ\,f(\vJ)$ specifies a valid galaxy
model of mass
\[\label{eq:Mint}
M=(2\pi)^3\int\d^3\vJ\,f(\vJ).
\]

The actions are defined by integrals
 \[\label{eq:defJ}
J_i={1\over2\pi}\oint_{\gamma_i}\d\vx\cdot\vv,
\]
 where $\gamma_i$ is a closed path in phase space. If we require that the
first action $J_r$ quantifies the extent of a star's radial excursions and
the third action $J_z$ quantifies the extent of its excursions either side of the
potential's equatorial plane, then the actions are  unambiguously
defined. What we here call $J_r$ is sometimes called $J_R$ or $J_u$, and what
we call $J_z$ is sometimes called $J_\vartheta$ or $J_v$, but no significance
attaches to these different notations. In a spherical potential
$J_z=L-|J_\phi|$, where $L$ is the magnitude of the angular momentum vector.

To obtain the observable properties of  a model defined by $f(\vJ)$, for example its density
distribution $\rho(\vx)=\int\d^3\vv\,f(\vJ)$ and its velocity dispersion tensor
$\sigma^2_{ij}(\vx)$, one has to be able to evaluate $\vJ(\vx,\vv)$ in an
arbitrary gravitational potential. Recently a number of techniques have been
developed for doing this \citep{Binney12a,SandersBinney14,SandersBinney15}.
Consequently, while the last word on action evaluation has likely not yet
been written, we now have algorithms that enable one to extract the
observables from a \df\ $f(\vJ)$ with reasonable accuracy.

DFs $f(\vJ)$ that depend on the phase-space coordinates only through the
actions were first used to model the disc of our Galaxy in an assumed
gravitational potential \citep{Binney10,Binney12b}.  Recently
\citet[][hereafter B14]{Binney2014} showed how to derive the self-consistent
gravitational potential that is implied by a given $f(\vJ)$ by exploring a
family of flattened, rotating models that he derived from the ``ergodic''
\df\ of the isochrone model: that is the \df\ $f(H)$ that depends on the
phase-space coordinates only through the Hamiltonian $H=\frac12
v^2+\Phi(\vx)$. \cite{Henon1960} derived the isochrone's ergodic \df, and in
the case of the isochrone potential explicit expressions are available for
$\vJ(\vx,\vv)$ and $H(\vJ)$ \citep{GerhardSaha}. Substituting $H(\vJ)$ in
$f(H)$ B14 obtained the \df\ $f(\vJ)$ of the isotropic isochrone model. In
this paper we present simple analytic functions $f(\vJ)$ that generate
nearly isotropic models of other widely used models, such as the
\cite{Hernquist1990}, \cite{Jaffe1983}, and \citet[][hereafter NFW]{NFW1996}
models. 

Once a \df\ of the form $f(\vJ)$ is available for a spherical, non-rotating
model, the procedure B14 used to flatten the isochrone sphere and to set it
rotating can be used to flatten and/or set rotating one's chosen model. So
\df s for spherical models in the form $f(\vJ)$ are valuable starting points
from which quite general axisymmetric models are readily constructed.

Galaxies are generally considered to consist of a number of components, such
as a disc, a bulge, and a dark halo, that cohabit a single gravitational
potential.  If we represent each component by a \df\ of the form $f(\vJ)$, it
is straightforward to find the gravitational potential in which they are all
in equilibrium \citep[e.g.,][]{Piffl14,Piffl15}.  An analogous composition
using \df s of the form $f(E,J_\phi,I_3)$ has never been achieved and may be
impossible, because when components are added, their potentials must be
added, and the energies of physically similar orbits in a given component are
quite different before and after we add in the potential of another
component. For example, the orbit on which a star sits at the centre of the
galaxy will have different energies before and after addition. If $E$ is used
as an argument of the \df, the change in $E$ will change the density of stars
on the given orbit, which is contrary to the fundamental idea of building up
the galaxy by adding components. By contrast, the actions of the orbit on
which a star sits at the galactic centre
vanish in any potential, and if a component is defined by $f(\vJ)$, it
contributes the same density of stars to this orbit regardless of the
external potential in which that component finds itself.
This fact is a major motivation for discovering what \df\
of the form $f(\vJ)$ is required to generate each component of a galaxy.

The \df\ of an isotropic spherical model must depend on the actions only via
the Hamiltonian $H(\vJ)$. The dependence of $f$ on $H$ is readily obtained
from the inversion formula of \cite{Eddington1916}, but an exact expression
for $H(\vJ)$ is only available for the isochrone potential and its limiting
cases, the harmonic oscillator and Kepler potentials. Our ignorance of
$H(\vJ)$ for potentials other than the isochrone amounts to a barrier to the
extension of B14's approach to model building.  One way to break through this
barrier is to devise numerical approximations to $H(\vJ)$ and some success
has been had in this direction by \cite{FermaniThesis} and
\cite{Williams+2014}. In this paper we pursue a slightly different strategy,
which is to develop simple algebraic expressions for \df s $f(\vJ)$ that
generate self-consistent models that closely resemble popular spherical
systems. We also show that a very simple form of $f(\vJ)$ generates a model
that is almost identical to the isochrone sphere and we give a useful
analytic expression for the radial action as a function of energy and angular
momentum for a Hernquist sphere.

The paper is organised as follows. In Section~\ref{sec:powerlaw} we use
analytic arguments to infer $f(\vJ)$ for scale-free models. These models are
not physically realisable as they stand, so in Section~\ref{sec:twopow} we
consider models that consist of two power-law sections joined at a break
radius. In Section~\ref{sec:CoresCuts} we extract realisable models from
scale-free models by the alternative strategy of adding a core to the system
and/or tidally truncating the model. Section~\ref{sec:conclude} sums up.

\section{Power-law models}
\label{sec:powerlaw}

Consider a gravitational potential that scales as a power of the distance
from the galactic centre, i.e.  $\Phi(\xi\bx)\propto\xi^a\Phi(\bx)$ with
$a\not= 0$: in the limit $a\to 0$ the gravitational potential tends to a
logarithmic potential, which is an interesting special case that we will
treat in Section \ref{sec:logarithmic}.

An orbit in a power-law potential has time-averaged kinetic and potential
energies, $K$ and $W$ respectively, that are related by the virial theorem:
$2K=a W$. The instantaneous total energy, given by the sum of the
instantaneous kinetic and potential energies, is conserved along the orbit
and consequently is given by
\begin{equation}
\label{eq:virial}
E=K+W=\left( \frac{a}{2} +1\right)W.
\end{equation}
In any power-law potential we need only to study orbits of one arbitrarily
chosen energy $E$ because each of these orbits can be rescaled to a similar
orbit at any given energy $E'$.  Indeed, if an orbit is rescaled by a spatial
factor, i.e., $\bx\to\bx'=\xi\bx$, then the orbit's total energy scales as
\begin{equation}
\label{eq:energy_scal}
E \to E'=\xi^a E,
\end{equation}
since obviously $W \to W'=\xi^a W$.  Further $v^2\propto K=\fracj12aW$, so
under rescaling $\vv \to \vv'=\xi^{a/2}\vv$. 

Given the scalings derived
above for $\vx$ and $\vv$ it follows that
\begin{equation}
\label{eq:action_scal}
\vJ \to \vJ'=\xi^{1+a/2}\vJ.
\end{equation}
Thus both the energy and the actions of an orbit that is rescaled by the
spatial factor $\xi$ are rescaled by powers of this factor.  

From equations \eqref{eq:energy_scal} and \eqref{eq:action_scal} we deduce
that the Hamiltonian is of the form
\begin{equation}
\label{eq:H(J)_pow1}
H(\vJ) = [h(\vJ)]^{a/(1+a/2)},
\end{equation}
where $h(\vJ)$ is a homogeneous function of degree one, i.e.
$h(\zeta\vJ)=\zeta h(\vJ)$ for every constant $\zeta$. In particular, $H$ is
itself a homogeneous function of the three actions of degree $a/(1+a/2)$.  It
is easy to check that equation \eqref{eq:H(J)_pow1} gives the correct
scalings $H\propto|\vJ|$ and $H\propto|\vJ|^{-2}$ for the harmonic oscillator
($a=2$) and Kepler $(a=-1$) potentials. \cite{Williams+2014} derive a closely
related result in which a specific form is proposed for $h(\vJ)$.

The homogeneous function $h$ is strongly constrained by the orbital
frequencies.  Indeed
\[\label{eq:freq_req}
{\Omega_i\over\Omega_j}={\pa H/\pa J_i\over\pa H/\pa J_j}
={{\pa h/\pa J_i}\over{\pa h/\pa J_j}}.
\]
 In a scale-free model the frequency ratio on the left is a homogeneous
function of degree zero, i.e., scale-independent, in agreement with the
right side. A natural choice for $h$ that we will use extensively is
\[\label{eq:my_h}
h(\vJ)=J_r+{\Omega_\phi(\vJ)\over\Omega_r(\vJ)}|J_\phi|
+{\Omega_z(\vJ)\over\Omega_r(\vJ)}J_z.
\]
 In a scale-free model this is homogeneous of degree one, as required.
Moreover so
long as the frequency ratios do not change rapidly within a surface of
constant energy in action space, the derivatives of $h$ satisfy equation
\eqref{eq:freq_req} to good precision.

In the definition \eqref{eq:my_h} of $h(\vJ)$ the modulus of the angular
momentum $J_\phi$ appears because we are concerned with the construction of
the part of the \df\ that is even in $J_\phi$. If we wish to set the model
rotating, we will add to this even part an odd part as discussed by B14.

Consider now the density distribution that generates a power-law potential.
In the spherical case
\footnote{In the non-spherical case
\[
\begin{split}
4\pi G\rho(\xi\vr)&={a+a^2\over \xi^2r^2}\Phi(\xi\vr)\\
&+{1\over\xi^2r^2\sin\theta}
{\pa\over\pa\theta}\left(\sin\theta{\pa\Phi(\xi\vr)\over\pa\theta}\right)
+{1\over \xi^2r^2\sin^2\theta}{\pa^2\Phi(\xi\vr)\over\pa\phi^2}\\
&=4\pi G\xi^{a-2}\rho(\vr).
\end{split}
\]
 Consequently $\rho$ and $\Phi$ have simple scalings with $r$ but they are
not necessarily functions of each other.} we have
\[
{\d\Phi(r)\over\d r}={\pa\Phi(\xi r)\over r\pa\xi}\bigg|_{\xi=1}={a\over
r}\Phi(r).
\]
 Hence
\[
4\pi G\rho={1\over r^2}{\d\over\d r}\left(r^2{\d\Phi\over\d r}\right)
={1\over r^2}{\d\over\d r}\left(ar\Phi\right)={a+a^2\over r^2}\Phi.
\]
 If $a=-1$ we recover the expected result $\rho=0$, but for $a\ne0$ we obtain
the polytropic relation  for index $n=1-2/a$ \citep[e.g][\S4.3.3a]{BT08}:
\[\label{eq:poly_rho_Phi}
\rho\propto|\Phi|^{1-2/a}.
\]
 From this relation it is easy to derive the
ergodic \df
\[\label{eq:EvDF}
f(E)\propto E^{-(4+a)/2a}
\]
 from Eddington's formula \citep[e.g.][]{Evans1994}. From equations
\eqref{eq:H(J)_pow1} and \eqref{eq:EvDF} it follows that the distribution
function of a power-law model is
 \begin{equation}
\label{eq:df_pow}
f(\vJ) = [h(\vJ)]^{-(4+a)/(2+a)}.
\end{equation}
The \df\ of a power-law model is itself a power-law of the three actions and the exponent is completely determined
by that of $\Phi(\vx)$.

\subsection{Logarithmic potentials}
\label{sec:logarithmic}

Now consider the limit $a\to0$ when the scaling of $\Phi$ becomes additive
\[
\Phi(\xi\vx)=\Phi(\vx)+\Vc^2\log(\xi),
\]
 where $\Vc$ is a constant that one can easily show is the circular speed.
Since galaxies have quite flat circular-speed curves, potentials of this form
are very useful.

The kinetic energy $K$ does not change on rescaling, while the potential
energy $W \to W'=W+\Vc^2\log(\xi)$, so
\begin{equation}
\label{eq:energy_scal_log}
E \to E'=E+\Vc^2\log(\xi).
\end{equation}
 The invariance of $K$ implies invariance of  $\vv$ under orbit rescaling,
so the scaling of the actions is
\begin{equation}
\label{eq:action_scal_log}
\vJ \to \vJ'=\xi\vJ=\exp\left( \frac{E'-E}{\Vc^2} \right)\vJ.
\end{equation}
 We now use each side of this equation as the argument of a homogeneous
function of degree one, $h(\vJ)$, and obtain 
\[
h(\vJ')=\exp\left( \frac{E'-E}{\Vc^2} \right)h(\vJ),
\]
 or on  rearrangement
\[
E'=E+\Vc^2\log[h(\vJ')/h(\vJ)].
\]
 Here $E'$ and $E$ are the energies of any two orbits whose actions $\vJ'$ and
$\vJ$ are proportional to each other. We can choose to make $\vJ$ an orbit
with vanishing energy, and we can choose  $h$ to be the homogeneous
function that satisfies $h(\vJ)=1$ as $\vJ$ moves over the
surface $E=0$ in action space. With these choices, we have
\begin{equation}
\label{eq:H(J)_log}
H(\vJ')=\Vc^2\log[h(\vJ')].
\end{equation}

The ergodic \df\ that self-consistently generates the spherical logarithmic
potential is well known to be 
\begin{equation}
\label{eq:erg_expdf}
f(H)= \exp\left(\frac{E_0-H}{\sigma^2}\right),
\end{equation}
where $\sigma^2=\Vc^2/2$ and $E_0$ is a constant \cite[e.g.][\S4.3.3b]{BT08}.
Using equation \eqref{eq:H(J)_log} it follows that the ergodic \df\ is
\begin{equation}
\label{eq:df_log}
f(\vJ)= \hbox{constant}\times[h(\vJ)]^{-2}.
\end{equation}
 This result  is consistent with the limit $a\to 0$ of equation
\eqref{eq:df_pow} for a power-law model.

Note that equation \eqref{eq:df_log} implies that the phase-space density
diverges as $\vJ\to0$. It follows that this \df\ unambiguously specifies the
singular isothermal sphere, in contrast to the \df\ \eqref{eq:erg_expdf},
from which one can derive both cored and singular isothermal spheres
\citep[e.g.][\S4.3.3b]{BT08}.  It is characteristic of \df s of the form
$f(\vJ)$ that they uniquely and transparently specify the phase-space density
both at the centre of the model ($\vJ=0$) and for marginally bound orbits
$(\vJ\to\infty)$. From a \df\ that depends on energy, by contrast, the
phase-space density at the centre of the model is implicitly specified by the
boundary condition adopted at $r=0$ when solving Poisson's equation for the
self-consistent potential.

The considerations of the last paragraph apply equally to the power-law \df s
\eqref{eq:df_pow}: although we used the standard form   \eqref{eq:EvDF} of
the energy-based \df\ of the polytropes to derive this \df, it implies
infinite phase-space density at the system's centre, so it is inconsistent
with familiar cored polytropes, such as the Plummer model.

\section{Two-power models}
\label{sec:twopow}

Any power-law model is problematic in the sense that the mass interior to
radius $r$ diverges as $r\to\infty$ if the density declines as $r^{-b}$ with
$b\le3$, and the mass outside radius $r$ diverges as $r\to0$ when $b\ge3$.
Hence there is no value of $b$ for which the model is physically reasonable at
both large and small $r$. One way we can address this problem is to assume
that $\rho$ scales as different powers of radius at small and large radii.
A widely used family of models of this type is given by the density profile
\[
\rho(r)={\rho_0\over(r/\rb)^\alpha(1+r/\rb)^{\beta-\alpha}},
\]
 where $\rb$ is the break radius \citep[e.g.,][]{BT08}. Three particular
cases of  importance are the \cite{Jaffe1983} model $(\alpha,\beta)=(2,4)$,
the \cite{Hernquist1990} model $(\alpha,\beta)=(1,4)$, which belong to the
family of \cite{Dehnen93} models $(\beta=4)$, and the NFW model
$(\alpha,\beta)=(1,3)$ \citep{NFW1996}. The ergodic \df s of the Jaffe and
Hernquist models are known analytic function, but that of the NFW model is
not. Our goal in this section is to find analytic functions $f(\vJ)$
that generate models that closely resemble these three classic models.

In the regime $r\ll \rb$ the mass $M(r)$ enclosed by the sphere of radius $r$
is $M\propto r^{3-\alpha}$, so the gravitational acceleration is $\d\Phi/\d
r\propto r^{1-\alpha}$ and thus the potential drop between radius $r$ and the
centre is
\[
\Phi(r)-\Phi(0)\propto r^{2-\alpha}\quad\hbox{or}\quad \log(r)\hbox{ when
}\alpha=2.
\]
 Setting $a=2-\alpha$ we can now employ the results we derived above for power-law potentials to
 conclude that 
\[
f(\vJ)=[h(\vJ)]^{-(6-\alpha)/(4-\alpha)}.
\]
 The Hernquist and NFW models both have $\alpha=1$ so we expect their \df s
 to have asymptotic behaviour
\[
f(\vJ)=[h(\vJ)]^{-5/3}\hbox{ as }|\vJ|\to0.
\] 
 A Jaffe model has $\alpha=2$, so the asymptotic behaviour of the Jaffe
model's \df\ as $\vJ\to0$ is given by equation \eqref{eq:df_log}.

Consider now the asymptotic behaviour of a two-power model as $r\to\infty$.
If the model has finite mass, the potential will asymptote to the Kepler
potential, $\Phi\propto r^{-1}$, so $\rho\propto|\Phi|^\beta$.  In the Kepler
regime the dependence of the Hamiltonian on the actions is
\citep[e.g.][eq.~3.226a]{BT08}
\[\label{eq:KeplerH}
H(\vJ)=[g(\vJ)]^{-2},
\]
 where $g(\vJ)$ is a homogeneous function of degree one.  Although $\rho$ is
a simple power of $|\Phi|$ we cannot employ the polytropic formula
\eqref{eq:EvDF}, because that rests on Poisson's equation, which does not
apply in this case: the model's envelope is a collection of test particles
that move in the Kepler potential generated by its core. We instead go
back to Eddington's formula
\[
f(\cE)\propto
{\d\over\d\cE}\int_0^{\cE}{\d\Psi\over\sqrt{\cE-\Psi}}{\d\rho\over\d\Psi},
\]
 where $\cE=-E$ and $\Psi=-\Phi$. From this formula it is easy to show that
 $\rho\propto\Psi^\beta$ implies 
\[
f(\cE)\propto\cE^{\beta-3/2}.
\]
 Combining this with equation \eqref{eq:KeplerH} we conclude that for
 $\beta>3$ the asymptotic behaviour of a double-power  \df\ is
\[
f(\vJ)=[g(\vJ)]^{-2\beta+3} \hbox{ as }|\vJ|\to\infty.
\]
 For the Jaffe and Hernquist models  $\beta=4$, so for these models
\[
f(\vJ)=[g(\vJ)]^{-5}\hbox{ as }|\vJ|\to\infty.
\]

Now that we have the asymptotic behaviour of $f$ in the limits of both small
and large $\vJ$, it is straightforward to devise a suitable form of the \df\
\[\label{eq:f_alpha_beta}
f(\vJ)={M_0 \over J_0^3}{[1+J_0/h(\vJ)]^{(6-\alpha)/(4-\alpha)}\over [1+g(\vJ)/J_0]^{2\beta-3}}.
\]
Here $M_0$ is a constant that has the dimensions of a mass and $J_0$
is a characteristic action. If the two homogeneous functions are normalised
such that $h(\vJ)\simeq g(\vJ)\simeq |\vJ|$, orbits that linger near the
break radius $\rb$ have $|\vJ|\simeq J_0$. These conditions ensure that $f$
tends to the required powers of $h$ and $g$ when $|\vJ|\ll J_0$ and $|\vJ|\gg
J_0$, respectively. 

We use different homogeneous functions for the regimes of small and large
$\vJ$ because the frequency ratios in these two regimes will differ. In the
Kepler regime, which is handled by $g$, all  frequencies are equal, so if
we require an isotropic model we choose
\[\label{eq:gJ}
g(\vJ)=J_r+|J_\phi|+J_z. 
\]
 In the regime of small $\vJ$, $\Omega_r>\Omega_\phi=\Omega_z$, and we take
$h$ to be of the form \eqref{eq:my_h} with a frequency ratio that is less
than unity.  Unfortunately, in this regime the
frequency ratio does vary over a surface of constant energy and an exactly
isotropic model cannot be constructed using constant ratios.
We simply use $\Omega_\phi/\Omega_r = \Omega_z/\Omega_r = 1/2$,
which are the frequency ratios of a  harmonic
oscillator.

The \df\ \eqref{eq:f_alpha_beta} is infinite on the orbit $\vJ = 0$ of a star
that is stationary at the model's centre.  \emph{Cuspy} models such as the
Hernquist, Jaffe and NFW models do have such centrally divergent \df s, while
in other \emph{cored} systems the phase space density reaches a finite
maximum. Cored systems will be treated in Section \ref{sec:CoresCuts}.

\begin{table}
\begin{center}
\caption{The ratio of the half-mass radius $\rh$ to the scale radius $r_0$,
defined by equation \eqref{def:r0}, for the $f(\vJ)$ Isochrone, $f(\vJ)$ Hernquist and
$f(\vJ)$ Jaffe models.  For comparison we list also the ratio $\rh/\rb$, where $\rb$
is the break radius, of the corresponding classical models.}

\begin{tabular}{lccc} \label{tab:models}
& Isochrone & Hernquist & Jaffe \\ \hline\hline
$\rh/r_0$ & $3.4$ & $2.42$ & $0.76$ \\
$\rh/\rb$ & $3.06$ & $2.41$ & $1$ \\ \hline
\end{tabular}
\end{center}
\end{table}

\subsection{Technicalities}

Here we touch on some technical issues that arise when one sets out to
recover the observable properties of a model from the \df\ that defines it.
The first step is to normalise the \df\ to the desired total mass by
evaluating the integral \eqref{eq:Mint}.  When the \df\ depends only on the
function $h(\vJ)$ defined by equation \eqref{eq:my_h} [i.e., the case
$g(\vJ)=h(\vJ)$] with the frequency
ratios $\omega\equiv\Omega_\phi/\Omega_r=\Omega_z/\Omega_r$ taken
to be constant, it is convenient to change coordinates from $(J_r,J_\phi,J_z)$
to $(J_r,L,J_z)$ and integrate out $J_z$, and then to change coordinates to
$(h,L)$ and integrate out $L$. Then one finds
\[\label{eq:MfromJa}
{M\over(2\pi)^3}=\int \d h\,f(h)\int_0^{h/\omega}\d L\,L
={1\over2\omega^2}\int_0^\infty\d h\,h^2f(h).
\]
 In the more general case, when $h(\vJ)\neq g(\vJ)$, the integral \eqref{eq:Mint}
cannot be reduced to one-dimension. Equation (\ref{eq:MfromJa}) can
be written
\[\label{eq:Mint_adim}
{M\over(2\pi)^3}=M_0 \int \d\vy {[1+1/h(\vy)]^{(6-\alpha)/(4-\alpha)}\over [1+g(\vy)]^{2\beta-3}},
\]
 where $\vy\equiv\vJ/J_0$. The integral in equation \eqref{eq:Mint_adim} is
dimensionless and depends only on the model's parameters $\alpha,\beta$ and
on the forms of the homogeneous functions $h$ and $g$. It can therefore be
computed at the outset. Then the value of $M_0$ can be set that ensures that
the model has whatever mass is required.

The physical scales of the models are determined by the action  scale
$J_0$ and by the mass scale $M_0$, so the natural length scale is
\[\label{def:r0}
r_0 \equiv {J_0^2 \over GM_0}.
\]
In following sections we will present $f(\vJ)$ analogues of three classic
models that have a finite mass: the Hernquist, Jaffe and isochrone models.
For our analogue models the top row of Table ~\ref{tab:models} gives the
ratio $r_{\rm h}/r_0$ of half-mass radius to the scale radius defined by
equation \eqref{def:r0}. The second row gives for the classical models the
ratio of $r_{\rm h}$ to the break radius, and we see that for the Hernquist
model $r_0=r_{\rm b}$ to good precision, while in the other two cases the
difference between $r_0$ and $r_{\rm b}$ is less than 25 per cent. 

Once $f(\vJ)$ has been normalised, we are able to determine the potential
$\Phi(\vx)$ that the model self-consistently generates by the iterative
procedure described by B14. 

\subsection{Worked Examples}
\subsubsection{The Hernquist model}
\label{sec:hernquist}

The \cite{Hernquist1990} model is an interesting example both because it is a
widely used model and because we can derive its ergodic \df\ as a function of the
actions for comparison with the $f(\vJ)$ model given by equation \eqref{eq:f_alpha_beta}
with $(\alpha,\beta)=(1,4)$, which hereafter we refer to as $f(\vJ)$ Hernquist model.

\begin{figure}
\begin{center}
\includegraphics[width=.9\hsize]{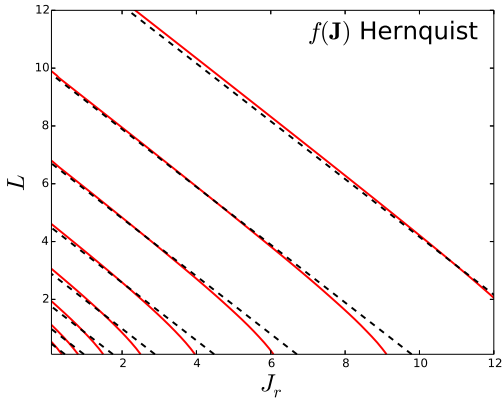}
\end{center}
\caption{The red full curves show  surfaces on which the \df\  of the
classical isotropic Hernquist sphere is constant in the
$(J_r,L)$ plane of action space, while the black dashed curves show surfaces on
which the corresponding $f(\vJ)$ distribution function is constant.}
\label{Hern_test}
\end{figure}

\begin{figure*}
\includegraphics[width=.33\hsize]{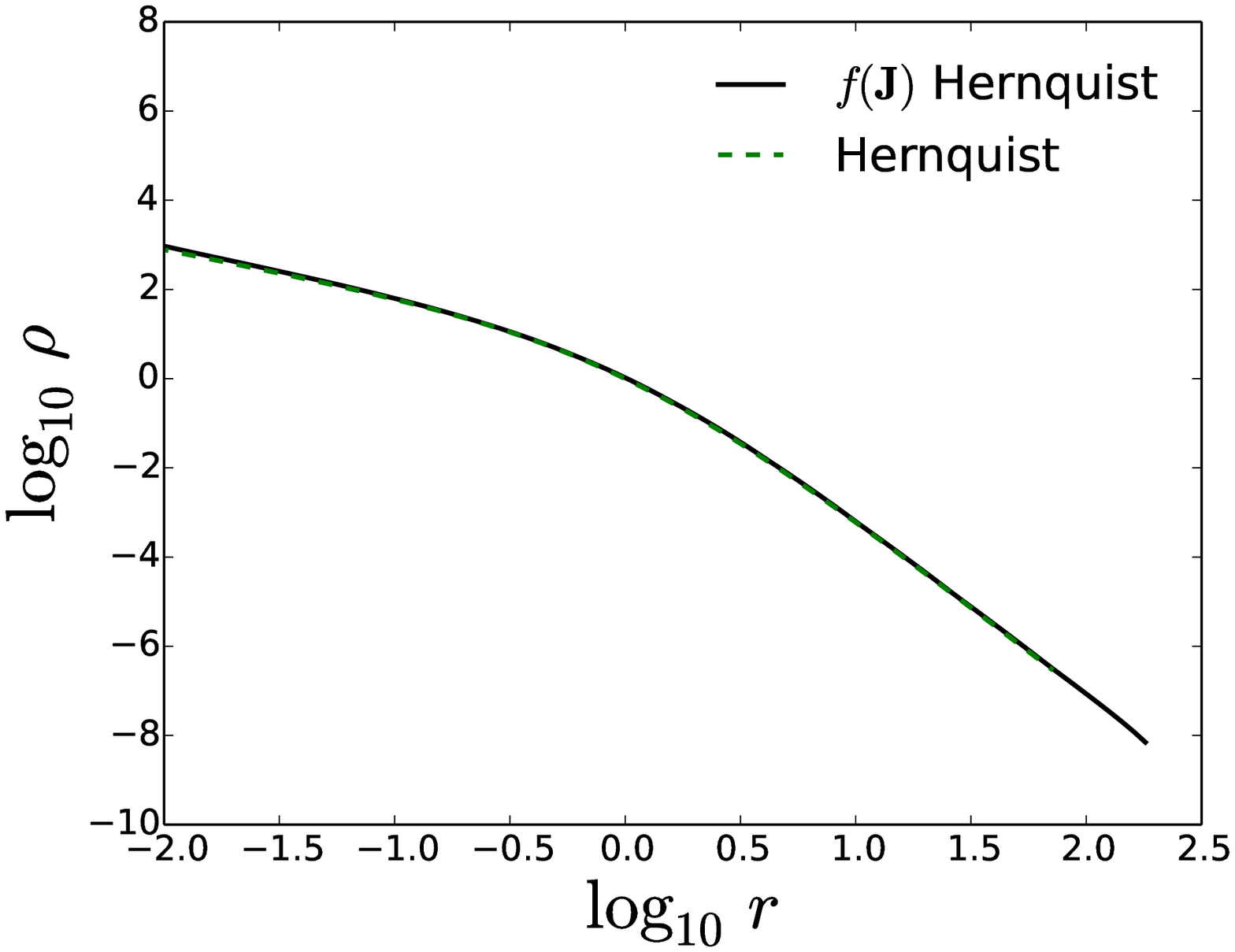}
\includegraphics[width=.33\hsize]{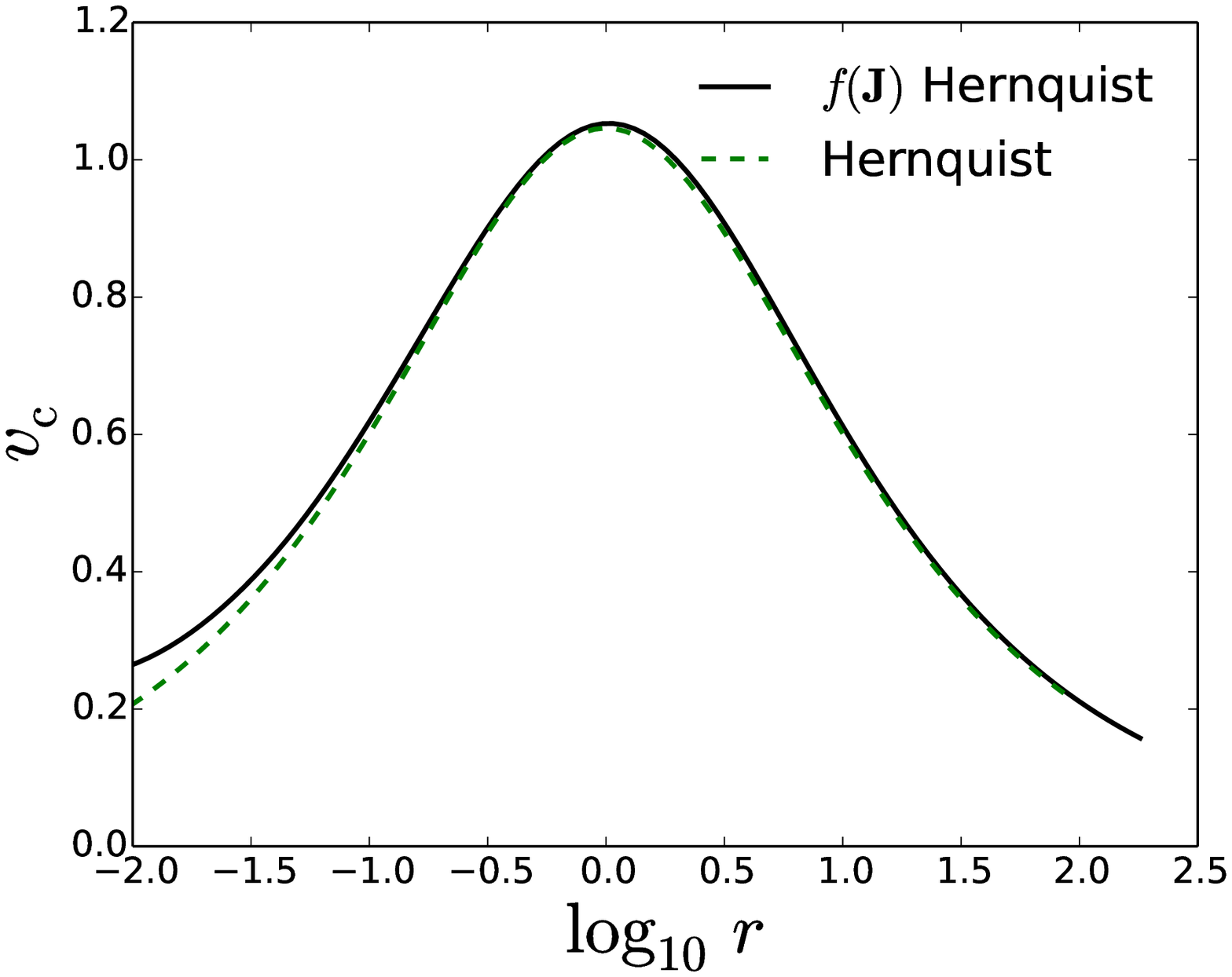} 
\includegraphics[width=.33\hsize]{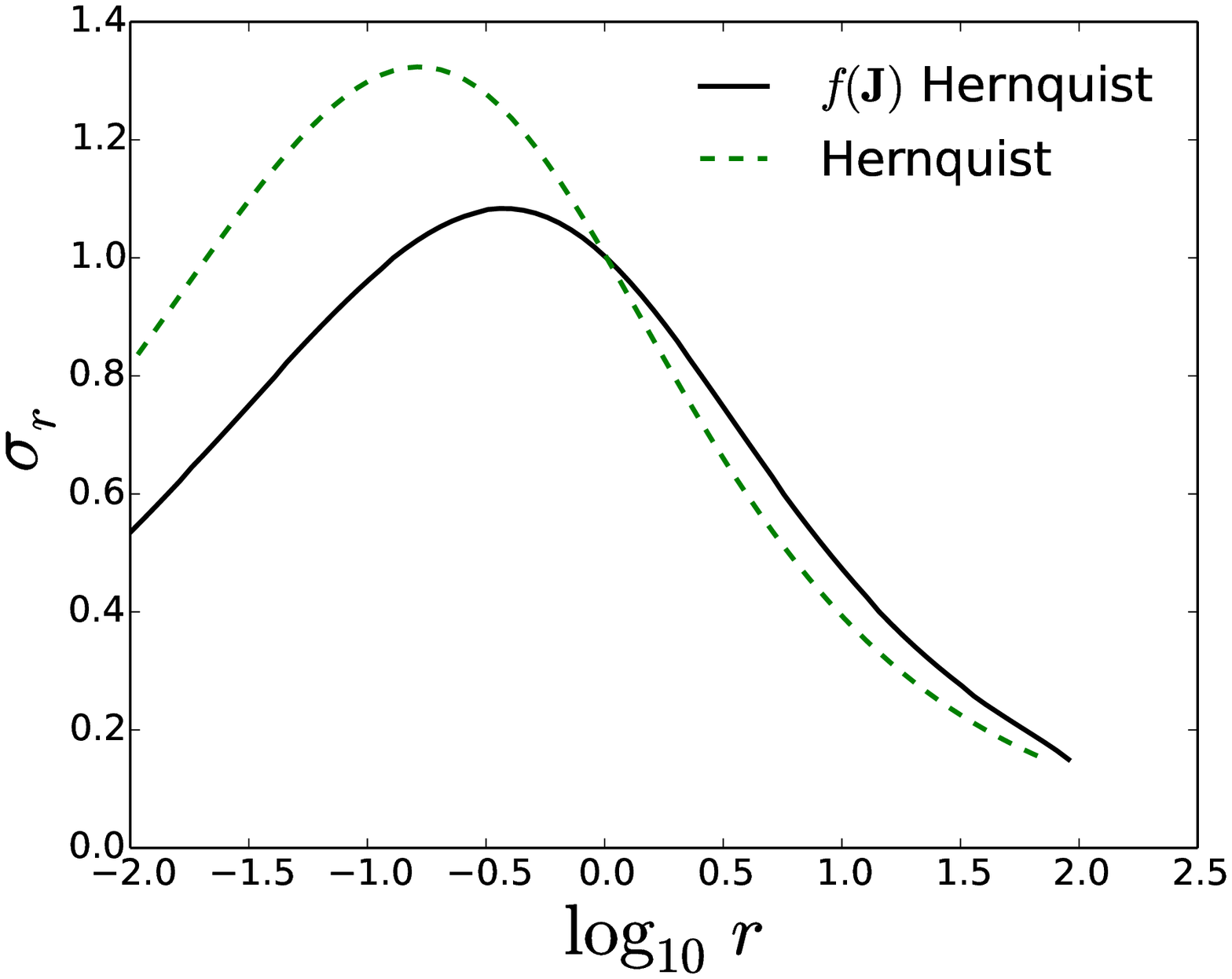}
\caption{Density (left-hand panel), circular velocity (central panel), and radial
velocity dispersion (left-hand panels) profiles for the classical isotropic Hernquist sphere
(normalized to $\rb$) and for the $f(\vJ)$ Hernquist model (normalized to $r_0$).}
\label{Hern_rad}
\end{figure*}

In Appendix \ref{app:JrHernquist} we derive an analytic expression for
$\Jr=\Jr(H,L)$ in the spherical Hernquist potential. By  numerically 
inverting this expression, we arrive at $H=H(\Jr,L)$ for the
Hernquist sphere. Combining this with the sphere's ergodic \df, which was
given already by \cite{Hernquist1990}, we have the exact $f=f[H(\vJ)]$.
In Fig.~\ref{Hern_test} we show surfaces in action space on which
this \df\  is constant together with surfaces on which \df\ of the
$f(\vJ)$ Hernquist model is constant.
The differences are small but apparent and arise because the
surfaces of constant energy are not exactly planar.

\begin{figure}
\begin{center}
\includegraphics[width=.8\hsize]{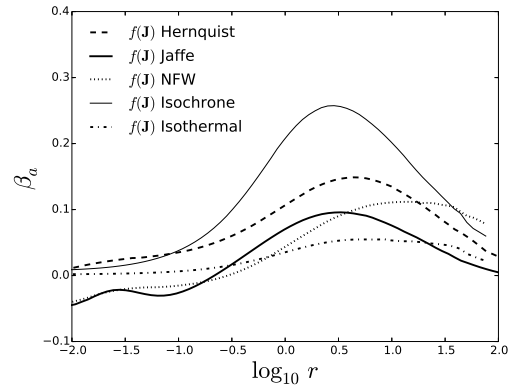}
\end{center}
\caption{Anisotropy profiles for $f(\vJ)$ Hernquist, $f(\vJ)$ Jaffe, $f(\vJ)$ NFW,
      $f(\vJ)$ isochrone and $f(\vJ)$ isothermal models. The profiles are normalized to
      $r_0$ (eq. ~\ref{def:r0}).
}
\label{fig:betaa}
\end{figure}

Fig.~\ref{Hern_rad} compares the radial profiles of
density, circular speed and radial component of velocity dispersion in the
exact isotropic model and in the $f(\vJ)$ Hernquist model.
The largest discrepancy is in the velocity dispersion and reflects the fact that
the model is significantly radially biased around $r_0$. The long-dashed
curve in Fig.~\ref{fig:betaa} shows
that the $f(\vJ)$ Hernquist model has a slight radial bias at
all radii by plotting the anisotropy parameter
\[\label{eq:betaa}
\beta_{\rm a}=1-{\sigma_\phi^2+\sigma_z^2\over2\sigma_r^2}.
\]
By virtue of the adopted form of $g$ (equation \ref{eq:gJ}), $\beta_{\rm a}
\to 0$ in the Keplerian regime.  Even though the potential is not harmonic at
the centre, still the model tends to isotropy also at small radii, which
justifies our simple choice for $h(\vJ)$.

\begin{figure*}
\includegraphics[width=.33\hsize]{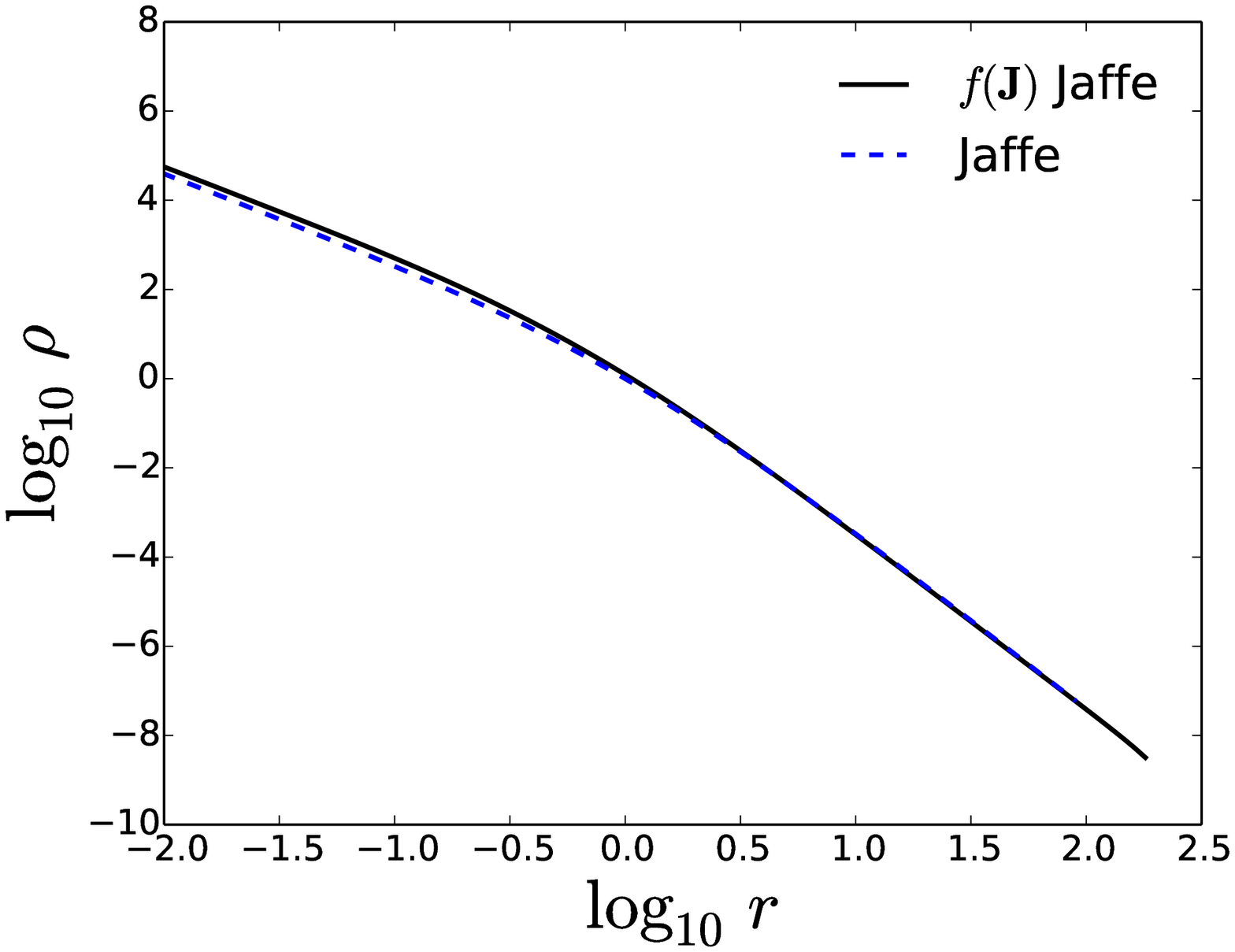}
\includegraphics[width=.33\hsize]{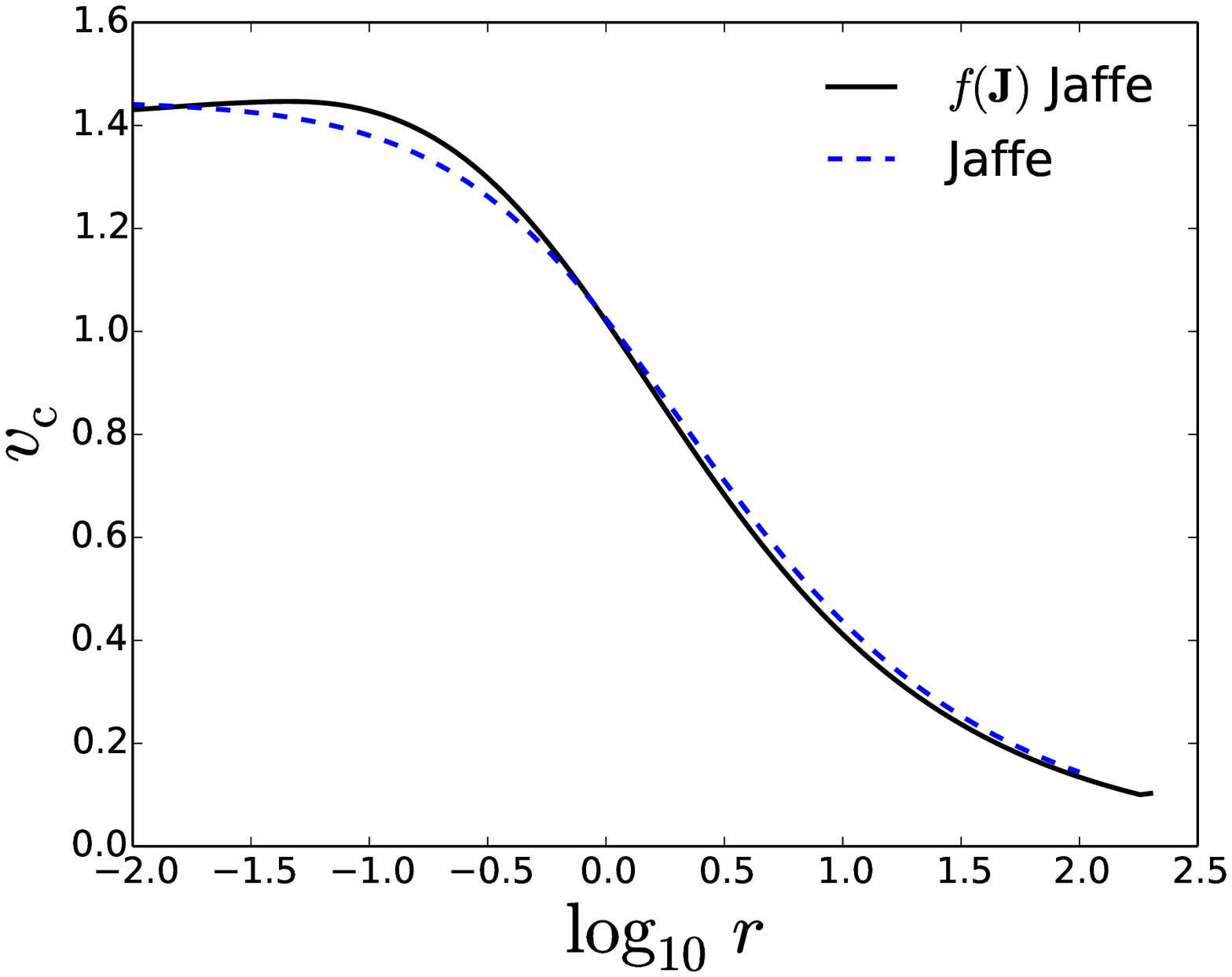} 
\includegraphics[width=.33\hsize]{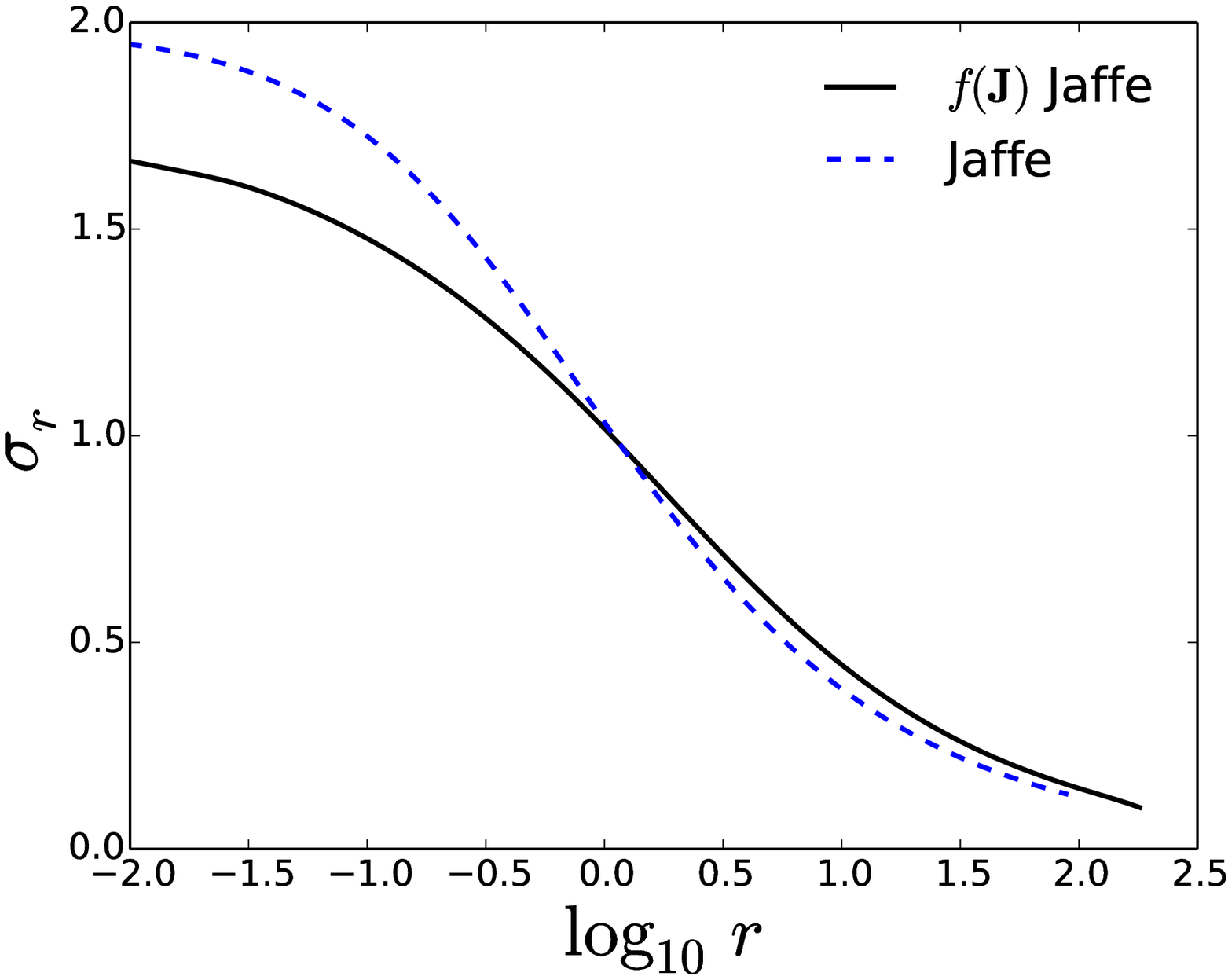}
\caption{Same as Fig. ~\ref{Hern_rad}, but for the classical isotropic Jaffe sphere and
for the $f(\vJ)$ Jaffe model.}
\label{fig:Jaffe_rad}
\end{figure*}

\subsubsection{The Jaffe model}
\label{sec:Jaffe}

The \cite{Jaffe1983} model behaves as Hernquist's at large radii, while
tending to $\rho\propto r^{-2}$ close to the centre.
Fig.~\ref{fig:Jaffe_rad} shows the radial profiles of the $f(\vJ)$ Jaffe
model defined by setting $(\alpha,\beta)=(2,4)$ in the \df\
\eqref{eq:f_alpha_beta}, and compares them with the classical isotropic model.
The discrepancies in $\sigma_r$ are due to the slight radial bias of the
$f(\vJ)$ model around $r_0$. The full curve in Fig.~\ref{fig:betaa} shows
that this bias actually quite  mild -- $|\beta_{\rm a}|<0.1$.

\subsubsection{NFW halo}
\label{sec:nfw}

\begin{figure*}
\includegraphics[width=.33\hsize]{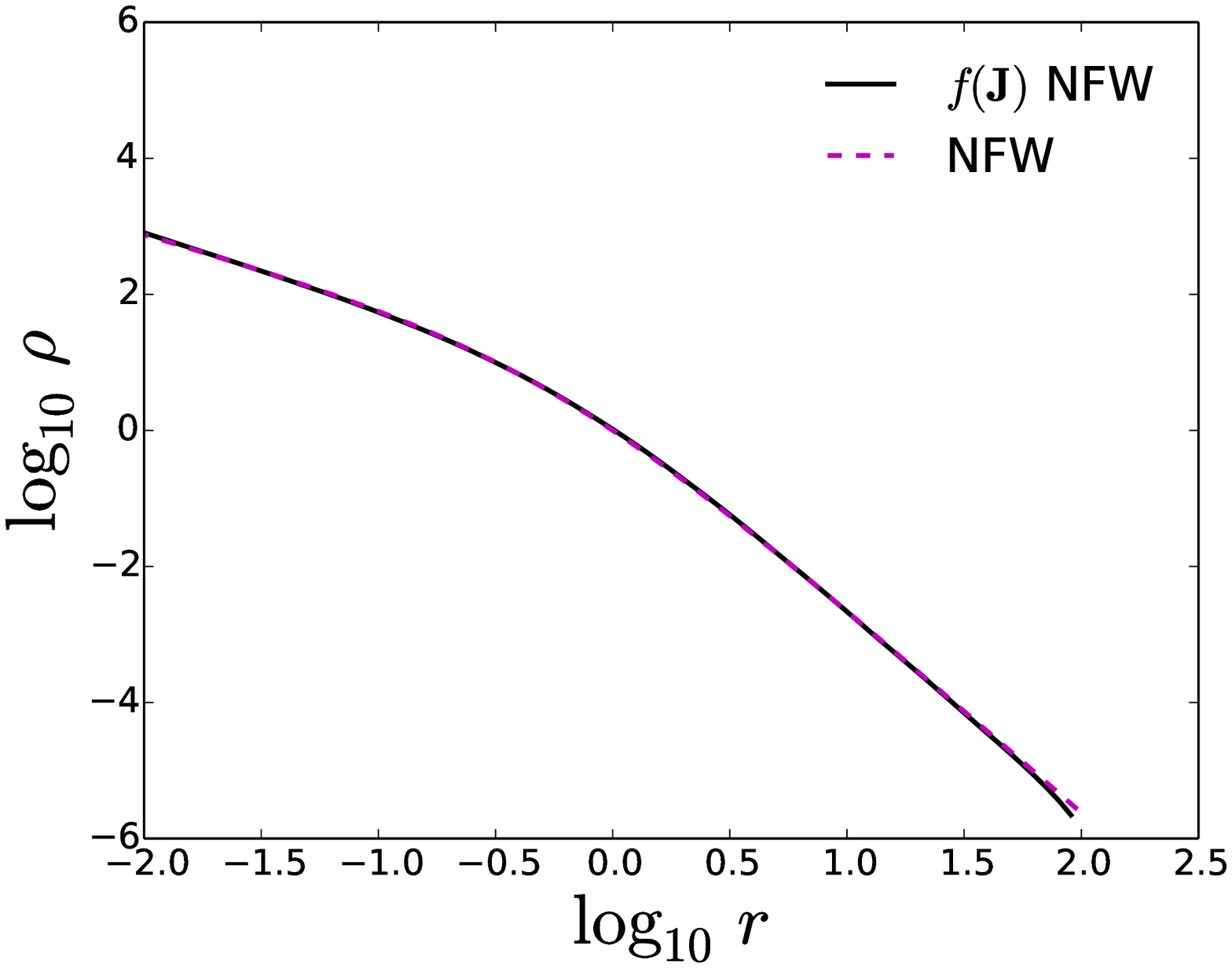}
\includegraphics[width=.33\hsize]{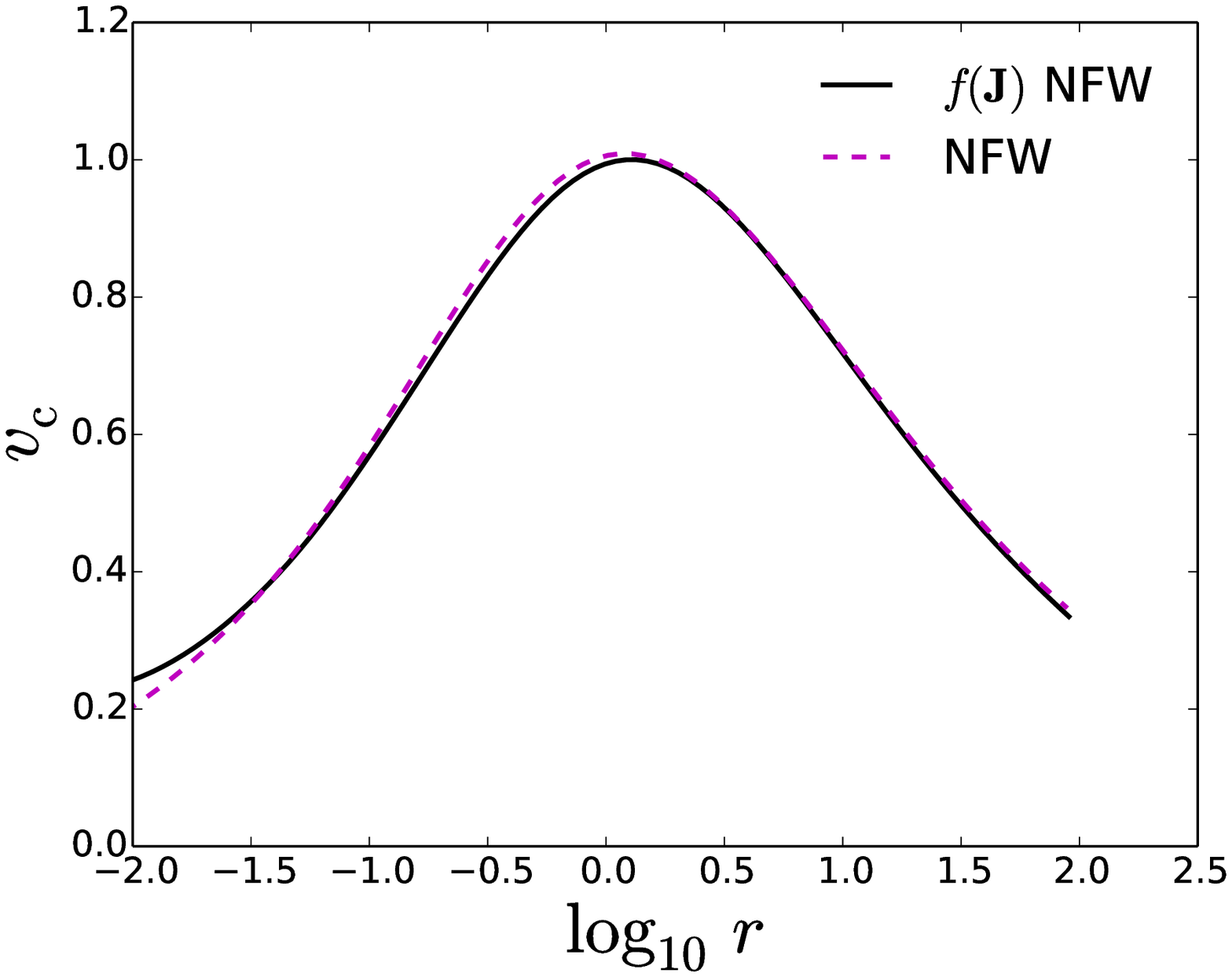}
\includegraphics[width=.33\hsize]{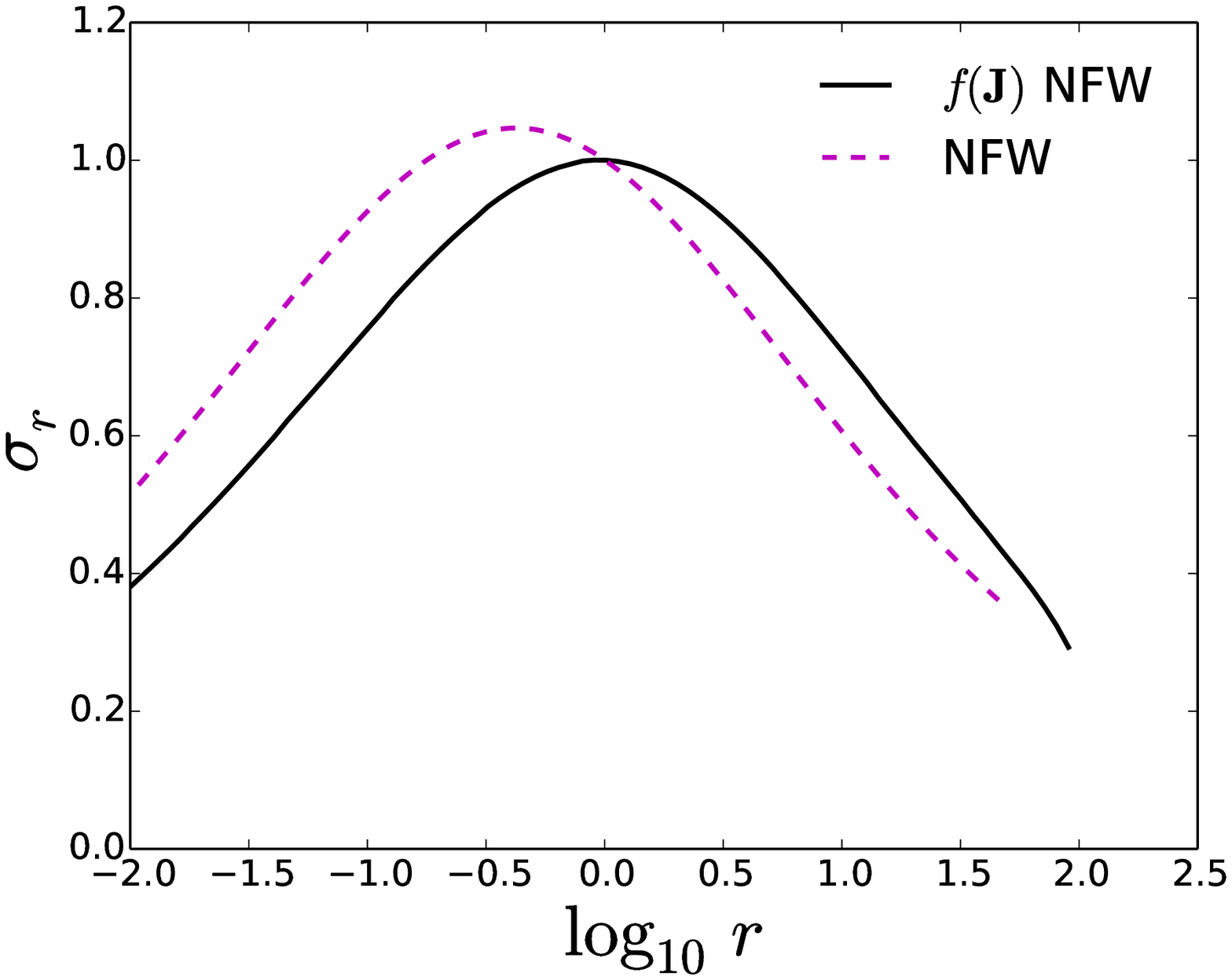}
\caption{Same as Fig. ~\ref{Hern_rad}, but for the classical isotropic NFW
sphere and for the $f(\vJ)$ NFW model defined by equation
\eqref{eq:almost_NFW}.} \label{NFW_rad}
\end{figure*}

The NFW model has $\beta=3$ with the consequence that its mass diverges
logarithmically as $r\to\infty$ and its potential is never Keplerian.
Consequently, the reasoning used to construct a \df\ above equation
\eqref{eq:f_alpha_beta} does not apply. If we nevertheless adopt equation
\eqref{eq:f_alpha_beta} with $(\alpha,\beta)=(1,3)$, we obtain a \df\ that
implies that as $J\to\infty$ the mass with actions less than $J$ diverges
like $\log J$. Asymptotically the circular speed of the standard NFW model is
\[
\Vc\sim \sqrt{{\log(1+r/r_0)\over r}},
\]
 so in this model the action of a circular orbit is $J_\phi\sim\sqrt{r\log
r}$. This shows that mass diverging like $\log J$ in action space
corresponds, to leading order, to divergence of the mass in real space like
$\log r$. Hence it is plausible that the \df\ \eqref{eq:f_alpha_beta} with
$(\alpha,\beta)=(1,3)$ generates a model similar to the NFW model. 

Computation of $\rho(r)$ for the $f(\vJ)$ model with $(\alpha,\beta)=(1,3)$
bears out this expectation. However the slope of the model's density profile at large
$r$ is slightly steeper than desired, and a better fit to the classical NFW
profile is obtained by adopting 
\[\label{eq:almost_NFW}
f(\vJ)={M_0 \over J_0^3}{[1+J_0/h(\vJ)]^{5/3}\over [1+g(\vJ)/J_0]^{2.9}}.
\]
 Fig.~\ref{NFW_rad} shows the radial profiles of the classical NFW model and
those of the model generated by the \df\ \eqref{eq:almost_NFW}, which we shall call
$f(\vJ)$ NFW model. The dotted curve in Fig.~\ref{fig:betaa} shows that this model
is mildly radially biased at radii larger than $r_0$ and it becomes very slightly
tangentially biased for $r<r_0$. These anisotropies account for the
difference between  the $\sigmar$ profiles of the $f(\vJ)$ and classical NFW
models.

\begin{figure*}
\includegraphics[width=.33\hsize]{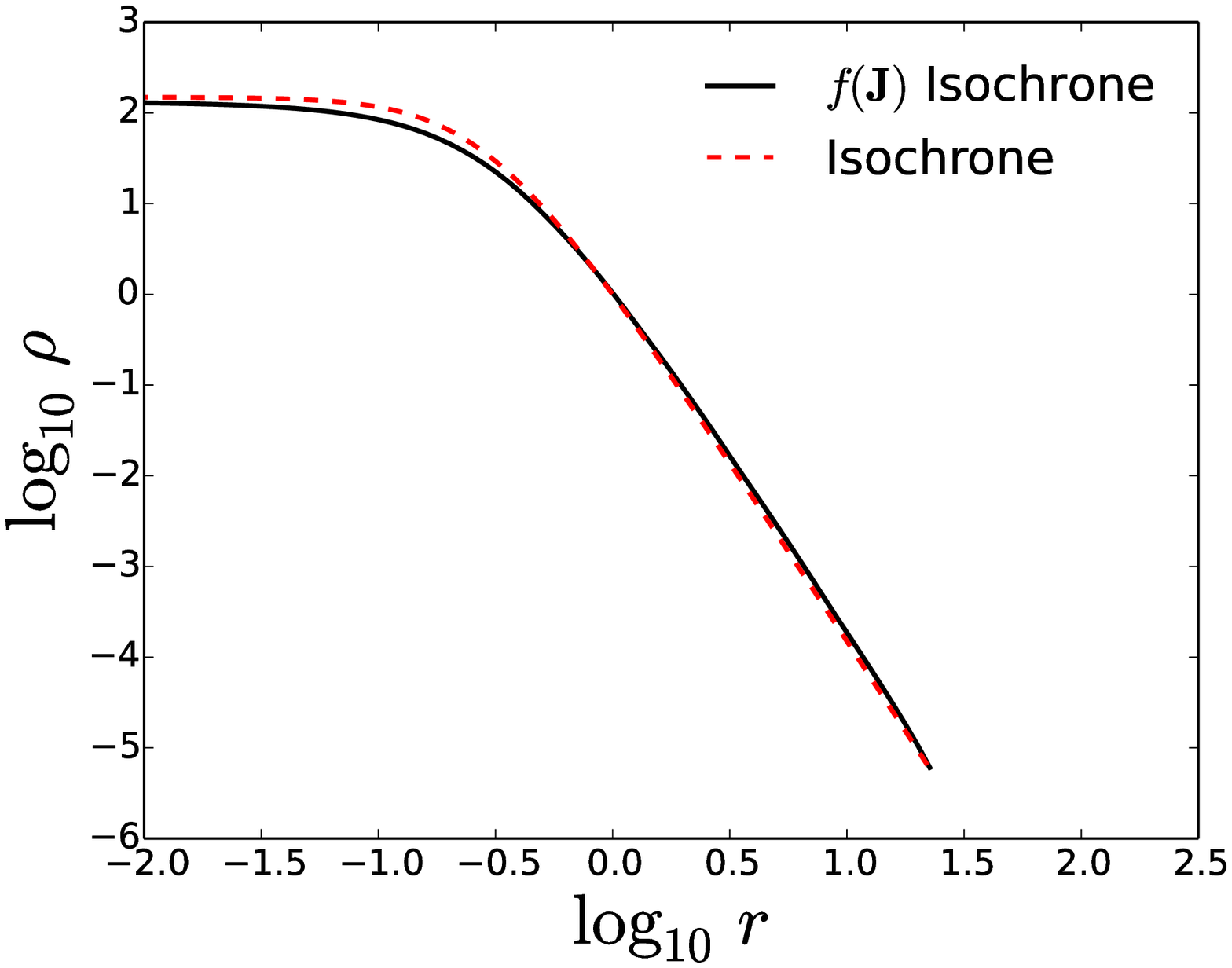}
\includegraphics[width=.33\hsize]{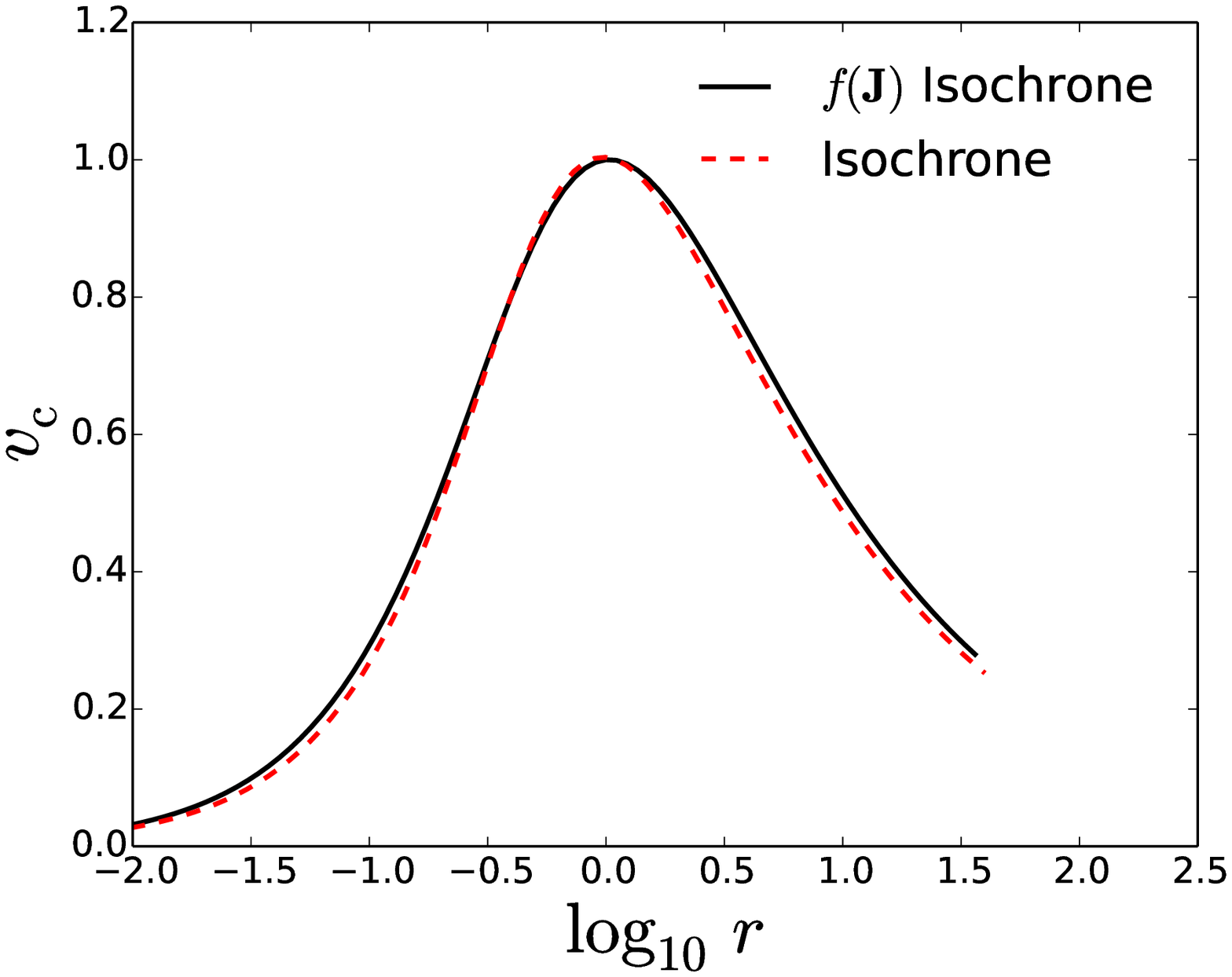} 
\includegraphics[width=.33\hsize]{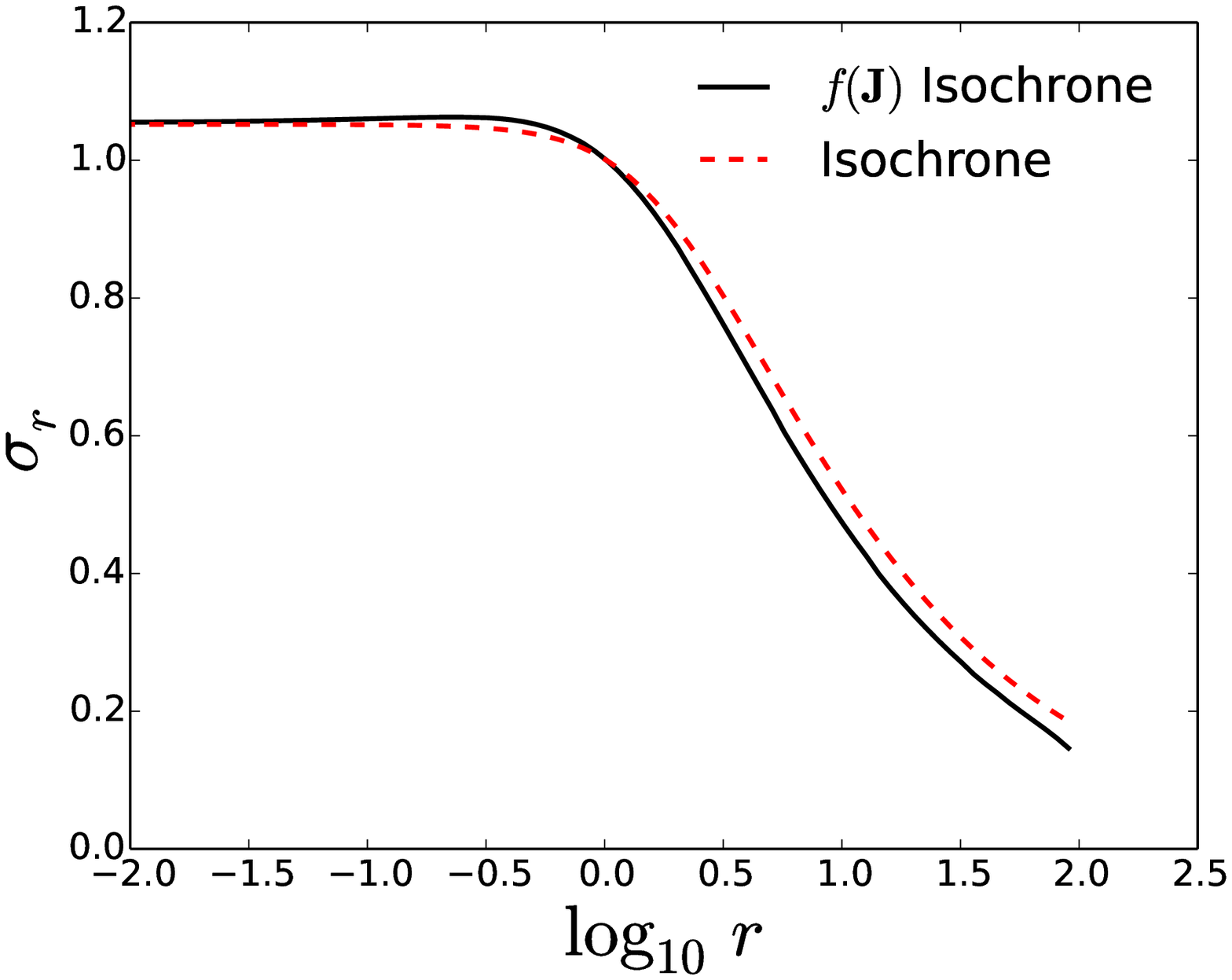}
\caption{Same as Fig. ~\ref{Hern_rad}, but for the classical isotropic isochrone sphere and
for the $f(\vJ)$ isochrone model.}
\label{fig:isoch_rad}
\end{figure*}
\begin{figure}
\begin{center}
\includegraphics[width=.8\hsize]{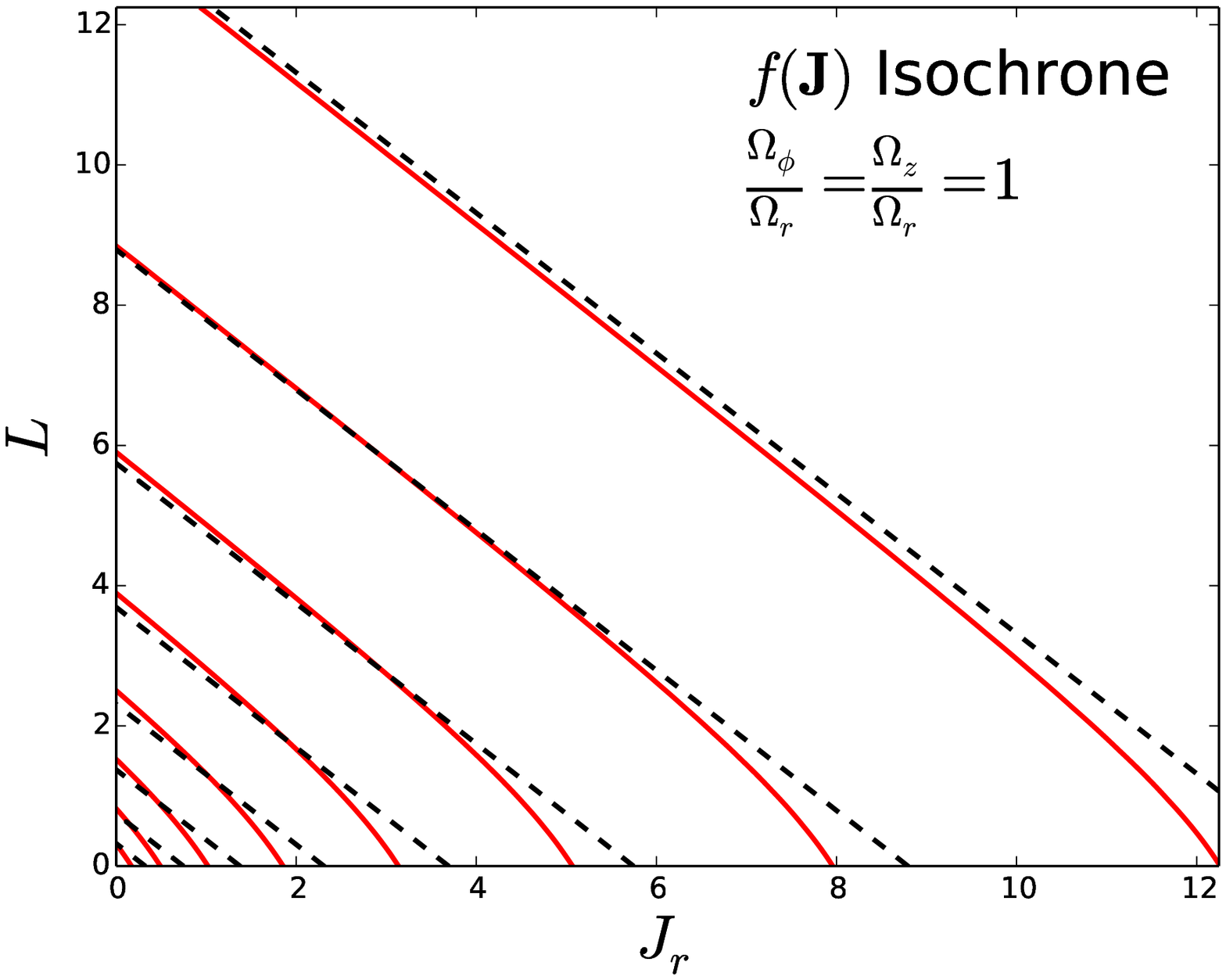}\\
\includegraphics[width=.8\hsize]{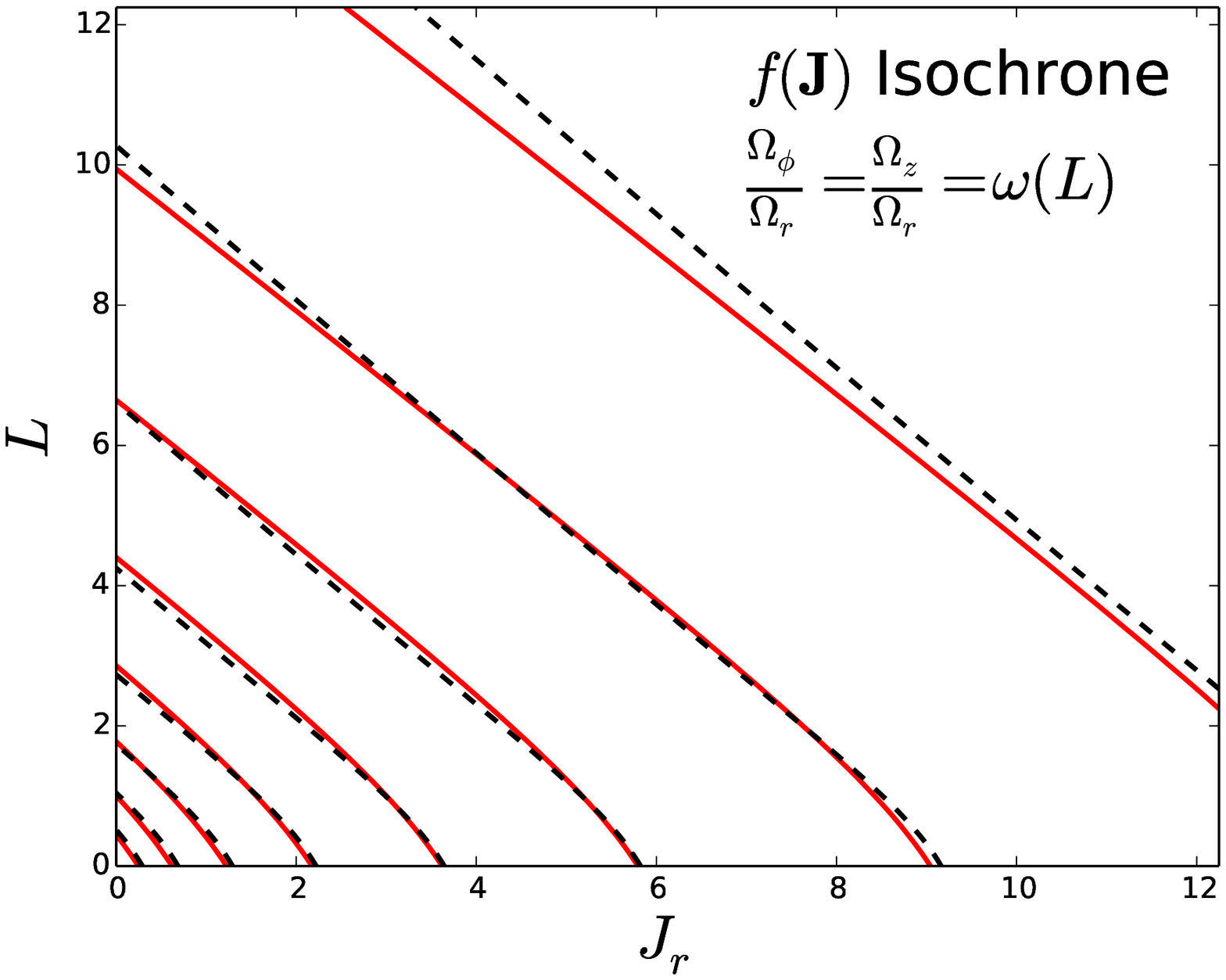}
\end{center}
\caption{Surfaces of constant $H(\vJ)$ (red) and of constant $f(\vJ)$ black
in action space for two $f(\vJ)$ isochrone models with different choice of the
function $g$ appearing in the \df\ \eqref{eq:coredDF}. In the upper panel $g$ is
$J_r+L$ whereas in the lower panel it is $J_r+(\Omega_\phi/\Omega_r)L$
where the frequency ratio is a function of $L$.} 
\label{fig:isoch_f_H}
\end{figure}

\section{Cores and cuts}\label{sec:CoresCuts}

In the last section we addressed the problematic nature of power-law
models -- that their mass diverges at either small or large radii -- by
introducing separate slopes of the dependence of $f$ on $\vJ$ at small and
large $\vJ$. The recovered models had central density cusps similar to those
of the Hernquist, Jaffe  and NFW models. If a homogeneous core is required, the
natural \df\ to adopt is
 \[\label{eq:coredDF}
f(\vJ)={M_0 \over J_0^3}{1\over[1+g(\vJ)/J_0]^{2\beta-3}},
\]
 for then the phase-space density has the finite value $M_0/J_0^3$ at the
centre of the model, and the asymptotic density profile is expected to be
$\rho\propto r^{-\beta}$.  For $\beta \leq 3$ the system has infinite mass,
so for these models we taper the \df\ by subtracting a constant from the
value given by equation \eqref{eq:coredDF}
 \[
f(\vJ) \mapsto f'(\vJ) = \max\left[0,f(\vJ) - f(\vJ_{\rm t})\right],
\]
 where $\vJ_{\rm t}$ is some large action, which defines a truncation radius
 \[\label{def:rtfromJt}
r_{\rm t}={|\vJ_{\rm t}|^2\over GM}.
\]

\subsection{Isochrone model}\label{sec:isochrone}

Fig.~\ref{fig:isoch_rad} compares the density profiles of the model equation
\eqref{eq:coredDF} generates for $\beta=4$ (black curves) with those of the
isochrone \citep{Henon1960}. The two models are extremely similar, so we
shall refer to the model generated by the \df\ \eqref{eq:coredDF} when $\beta=4$ as
the $f(\vJ)$ isochrone model. The density profiles of the two models are
essentially identical, but at $r \simeq r_0$ $\sigma_r$ is slightly smaller
in the $f(\vJ)$ isochrone than in the classical isochrone because the
$f(\vJ)$ isochrone is mildly radially biased near $r_0$ -- the thin full
curve in Fig.~\ref{fig:betaa} shows $\beta_{\rm a}(r)$ for this model. It
is non-zero because in action space surfaces of $f(\vJ)$ do not quite
coincide with surfaces of constant $H(\vJ)$, as the upper panel of
Fig.~\ref{fig:isoch_f_H} shows by plotting contours of $f$ and $H$.  For the
isochrone potential we have an analytic expression for the frequency ratio
$\Omega_\phi/\Omega_r$ as a function of $L$.  The lower panel of
Fig.~\ref{fig:isoch_f_H} shows that the constant-energy and constant-\df\
contours are more closely aligned when the argument of the homogeneous
function uses the exact frequency ratio.

Given that the exact \df\ of the isochrone is a complicated function of
$\vJ$, it is astonishing that the trivial \df\ \eqref{eq:coredDF} provides
such a good approximation to it.

\begin{figure*}
\includegraphics[width=.33\hsize]{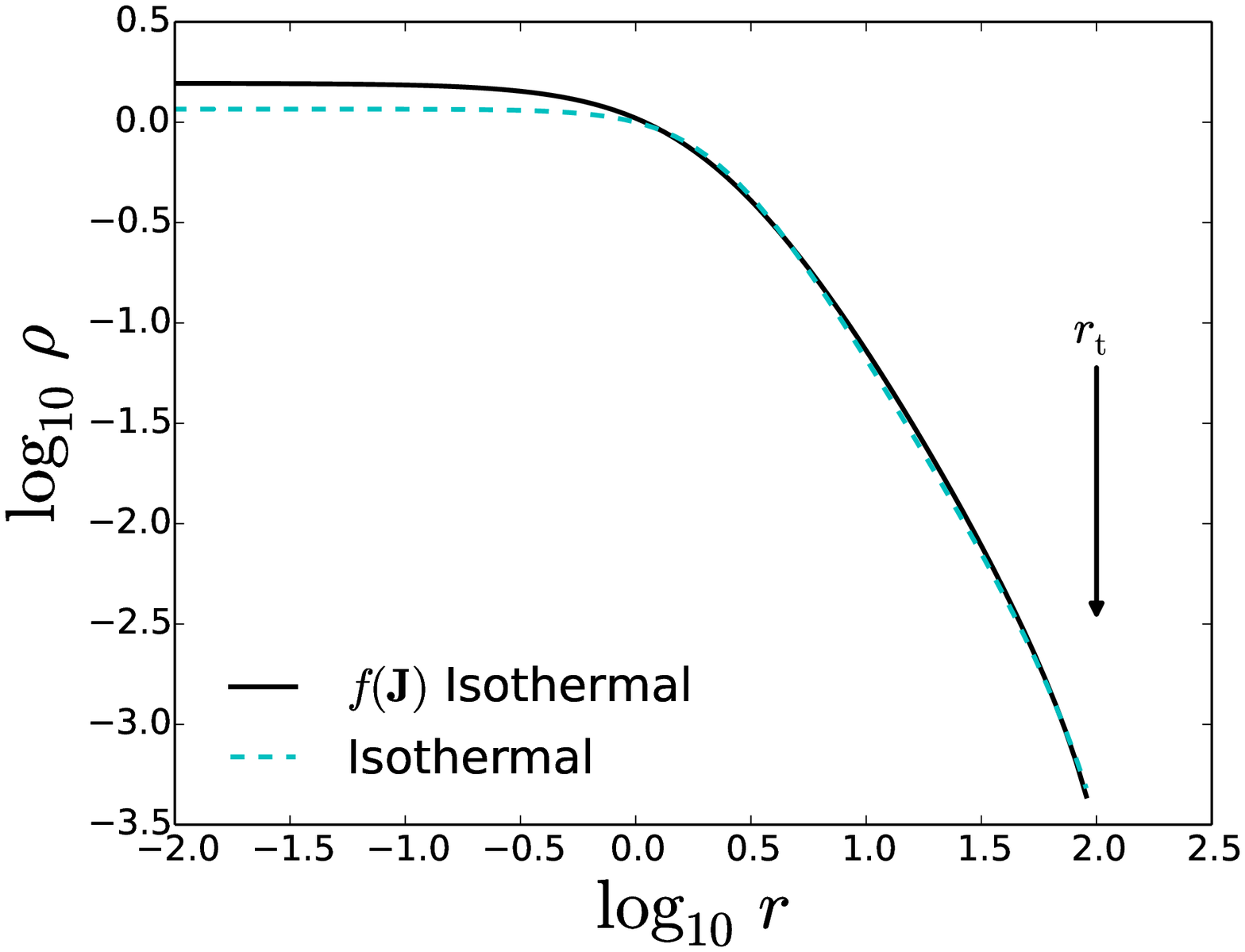}
\includegraphics[width=.33\hsize]{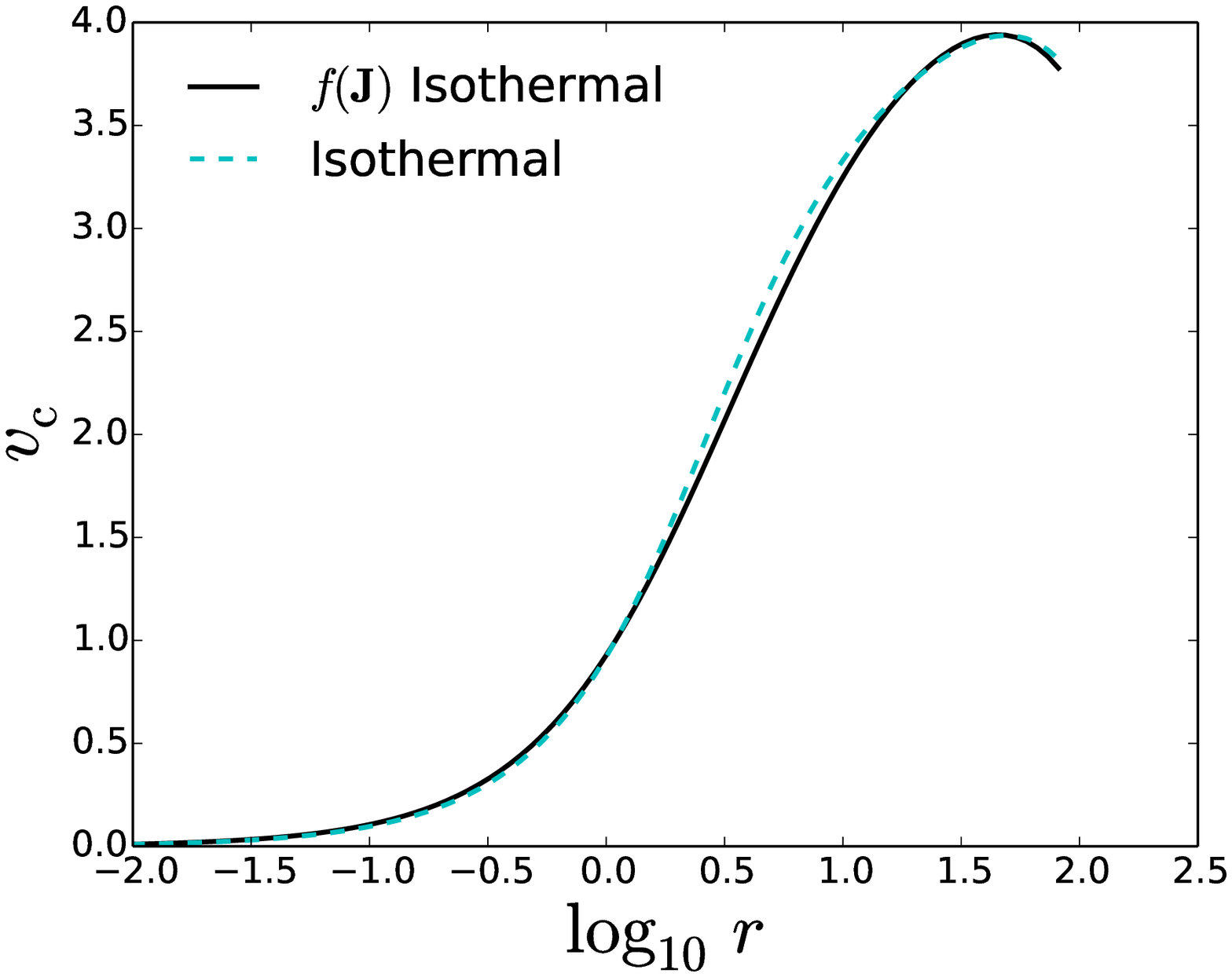} 
\includegraphics[width=.33\hsize]{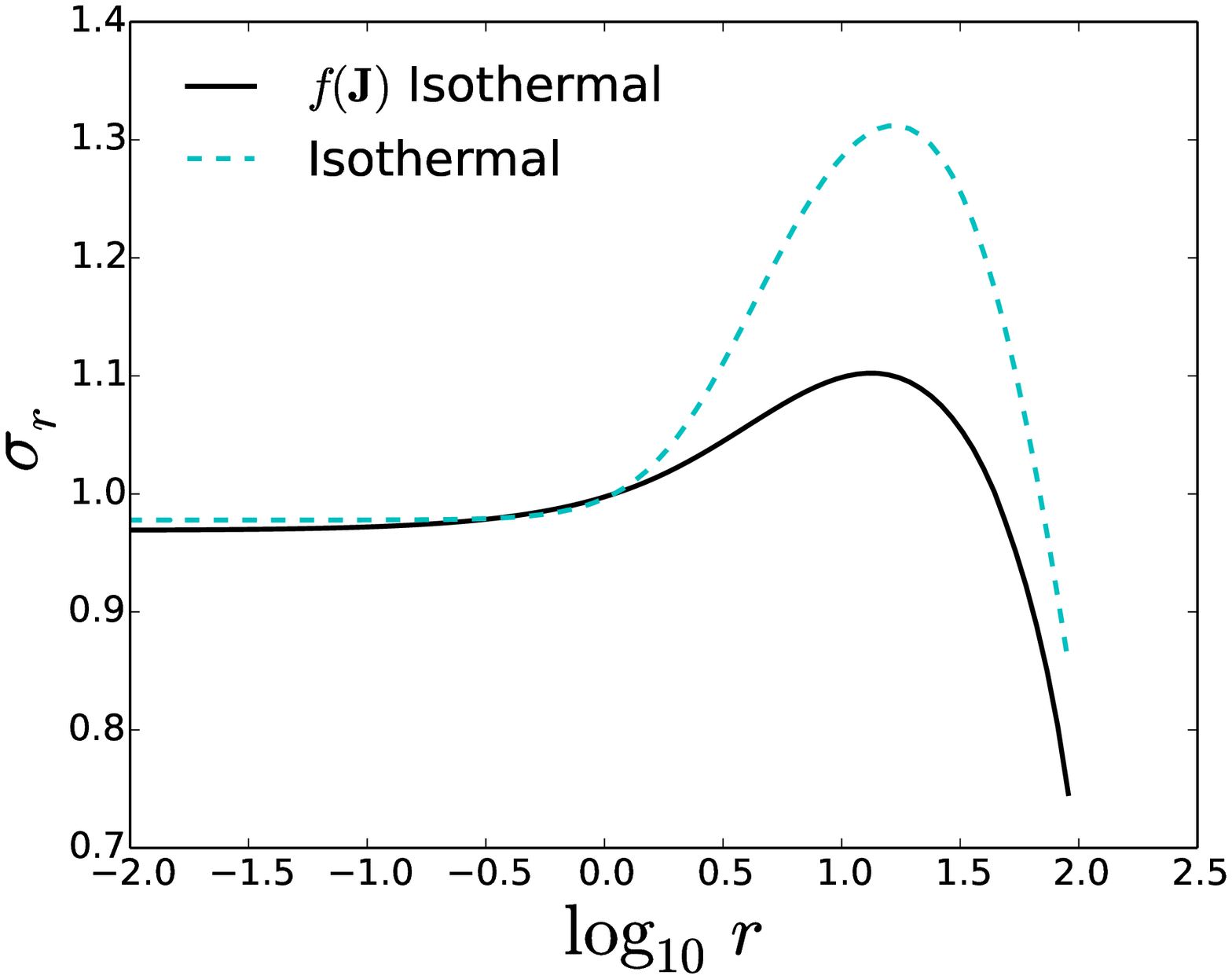}
\caption{Same as Fig. ~\ref{Hern_rad}, but for the truncated isotropic cored
isothermal sphere (equation \ref{eq:isoth_cored_trunc}) and for the $f(\vJ)$
truncated isothermal model. We show the location of the truncation radius
defined by equation \eqref{def:rtfromJt}.}
\label{isoth_rad}
\end{figure*}

\subsection{Cored isothermal sphere}\label{sec:isothermal}

In Section~\ref{sec:logarithmic} we derived an approximation
\eqref{eq:df_log} to the \df\ of the singular isothermal sphere. Here we
modify this model into one that is numerically tractable by (i) adding a
core, and (ii) tapering its density at large radii so the model's mass
becomes finite. Then the \df\ is
 \[\label{eq:cut_coredDF}
f(\vJ)={M_0 \over J_0^3}\max\left(0,[1+J_0/h(\vJ)]^2-{[1+J_0/h(\vJ_{\rm t})]^2}\right),
\]
 where $h(\vJ)$ is given by equation \eqref{eq:my_h} with both frequency
ratios set to $1/\surd2$ and $\vJ_{\rm t}=(0,\Vc r_{\rm t},0)$. As
full curves in Fig.~\ref{isoth_rad} show, this \df\ generates a model that has a core that
extends to $r_0$ and a density profile that plunges to zero near the
truncation radius $r_{\rm t}$.  The short-dashed curve in
Fig.~\ref{fig:betaa} shows the model's anisotropy parameter
$\beta_{\rm a}$, which is always small ($|\beta_{\rm a}|<0.04$). 

An ergodic model with a simple functional form of $\rho(r)$ to which we can
compare  our $f(\vJ)$ model has
\[\label{eq:isoth_cored_trunc}
\rho(r)={\Vc^2\over2\pi G(r^2+\rb^2)} \e^{-r^2/r_{\rm t}^2}.
\]
The dashed curve in the left panel-hand of Fig.~\ref{isoth_rad} shows that
the model defined by the \df\ \eqref{eq:isoth_cored_trunc} provides an excellent fit
to the density profile of our $f(\vJ)$ model. Curiously, in the $f(\vJ)$
model $\sigma_r(r)$ is more nearly constant within $r_{\rm t}$ than in either
of the models with analytic density profiles. The dashed curve in the
right-hand panel of Fig.~\ref{isoth_rad} shows that the model defined by
the \df\ \eqref{eq:isoth_cored_trunc} has a significantly deeper central depression in
$\sigma_r$ than the $f(\vJ)$ model.

\section{Conclusions}\label{sec:conclude}

Studies of both our own and external galaxies will benefit from the
availability of a flexible array of {\it dynamical} models of galactic
components such as disc, bulge and dark halo. The construction of general
models of this type is rather straightforward when one decides to start from
an expression for the component's \df\ as a function of the action integrals
$J_i$. In this paper we have illustrated this fact by deriving simple
analytic forms for \df s that self-consistently generate models that closely
resemble the isochrone, Hernquist, Jaffe, NFW and truncated isothermal
models. In previous papers \cite{Binney10,Binney12b} has given simple
analytic \df s that provide excellent fits to the structure of the Galactic
disc, so now \df s are available for all commonly occurring galactic
components.

Our models are tailored to minimise velocity anisotropy at both small and
large radii. In all of them the anisotropy parameter $\beta_{\rm a}$ peaks
at intermediate radii. The peak is by far sharpest in the $f(\vJ)$
isochrone, but even in this model $\beta_{\rm a}$ stays below $0.25$.

Our presentation has been elementary in the sense that we have confined
ourselves to spherical, almost isotropic components that live in isolation.
However, B14 showed that given a near-ergodic \df\ $f(\vJ)$ of a component
such as those presented here, it is trivial to modify it so it generates a
system that is flattened by velocity anisotropy, or by rotation, or by a
combination of the two.  Equally important, when the \df\ of an individual
component is given as $f(\vJ)$, it is straightforward to add components. Such
addition was exploited by \cite{Piffl14} in a study of the contribution
of dark matter to the gravitational force on the Sun: in that study
the models fitted to data comprised a sum of \df s $f(\vJ)$ for the disc and
the stellar halo. The dark halo was assigned a density
distribution rather than a \df, but \cite{Piffl15} represent the dark halo by
the $f(\vJ)$ NFW model, making the Galaxy a completely self-consistent
object. A key point for such work is that the mass of each component can be
specified at the outset.

Our approach has several points of contact with that of \cite{Williams+2014}
and \cite{Evans+2014}, who derive approximations to $H(\vJ)$ for models that
are defined by \df s of the form $f(E,L)$. In particular, they show that
for their models better approximations to the iso-energy surfaces in   action
space can be obtained if one's homogeneous function has as its argument the
sum of a linear function of the actions, as used here, and a small term
$\epsilon\sqrt{LJ_r}$. We expect that the anisotropy of our models could be
enhanced by adding such a term.

In addition to assisting in the dynamical interpretation of observations of
galaxies, the models that the present work makes possible could provide
useful initial conditions for N-body simulations. The first step would be the
construction of a self-consistent galaxy model from a judiciously chosen \df.
Then one could Monte-Carlo sample the action space using the \df\ as the
sampling density, and torus mapping \citep[e.g.][]{BinneyM11} could be used to
generate an orbital torus at each of the selected actions. Finally some
number $n$
of initial conditions $(\vx,\vv)$ would be selected on each torus, uniformly
space in the angles $\theta_i$. The resulting simulation would be in
equilibrium to whatever precision had been used in the solution of Poisson's
equation, and it would experience a ``cold start'' \citep{Sellwood87}.
Moreover, given that it would be possible to evaluate the original \df\ at
any phase-space point, the model would lend itself to the method of
perturbation particles \citep{Leeuwin93} in which the simulation particles
represent the difference between a dynamically evolving model and an
underlying equilibrium rather than the whole model. This method has been
little used in the past on account of the lack of interesting models with
known \df s, which is precisely the need that we have here supplied.

\section*{Acknowledgements}

LP is pleased to thank the Rudolf Peierls Centre for Theoretical Physics
in Oxford for the warm hospitality during an early phase of this work.
JB is supported by the European Research
Council under the European Union's Seventh Framework Programme
(FP7/2007-2013)/ERC grant agreement no.\ 321067, and by the UK Science
Technology through grant ST/K00106X/1.
LC and CN are partly supported by PRIN MIUR 2010-2011, project
``The Chemical and Dynamical Evolution of the Milky Way and Local Group
Galaxies'', prot. 2010LY5N2T.

\appendix

\section{Analytical expression for the radial action in the Hernquist sphere}
\label{app:JrHernquist}

The radial action is defined as
\begin{equation}
\label{ap:def_Jr}
\Jr = \frac{1}{2\pi} \oint p_r\d r 
= {1\over\pi}\int_{r_1}^{r_2}\de r\, \sqrt{2E-2\Phi(r)-\frac{L^2}{r^2}},
\end{equation}
where $\Phi(r) = -GM/(r+\rb)$ and $r_1,r_2$ are the pericentric and apocentric
radii for the given energy $E$ and angular momentum $L$, i.e., the two
roots of the integrand in equation \eqref{ap:def_Jr}.  Introducing the dimensionless
quantities $s \equiv r/\rb$, $\mE \equiv -E\rb/GM$ and $l = L/\sqrt{2GM\rb}$,
equation \eqref{ap:def_Jr} can be rewritten
get
\begin{equation}
\Jr = \frac{\sqrt{2GM\rb}}{\pi} \int_{s_1}^{s_2}\d s\, 
\sqrt{-\mE+\Psi(s)-\frac{l^2}{s^2}},
\end{equation}
where $\Psi(s)\equiv 1/(1+s)$ is the relative dimensionless potential.  We
now  change the integration variable
variable from $s = (1-\Psi)/\Psi$ to $\Psi$ \citep{Ciotti1996}, and have
\begin{equation}
\label{ap:Jrint}
\ds \Jr = \frac{\sqrt{2GM\rb}}{\pi} \int_{\Psi_2}^{\Psi_1}\d\Psi\,
 \frac{\sqrt{\mP(\Psi)}}{(1-\Psi)\Psi^2},
\end{equation}
where $\Psi_1\equiv\Psi(s_1)$, $\Psi_2\equiv\Psi(s_2)$ and
\begin{equation}
\label{ap:P1}
\mP(\Psi) = -\mE (1-\Psi)^2 + \Psi(1-\Psi)^2 -l^2\Psi^2
\end{equation}
is a cubic in $\Psi$, the roots of which can be found by standard methods 
\citep[e.g.,][]{Dickson1914}.
$\Psi_1, \Psi_2$ are two roots in the physical range
$0\leq\Psi\leq 1$. Let $A$ be the third real root, so
\begin{equation}
\label{ap:P2}
\mP(\Psi) = (\Psi_1-\Psi)(\Psi-\Psi_2)(A-\Psi).
\end{equation}
While it is physically obvious that two of the three real solutions of equation
\eqref{ap:P1} are in the range $(0,1)$ and the remaining one is outside $(A>1)$,
we remark that the same conclusion can be reached by purely algebraic arguments
by using the Routh-Hurwitz theorem \citep[see e.g.,][]{Gant59}.  By evaluating
equations \eqref{ap:P1} and \eqref{ap:P2} at $\Psi=0$ one gets
$A=\mE/\Psi_1\Psi_2 > 0$.
By splitting into its partial fractions the integrand in equation \eqref{ap:Jrint}, it is possible to
express the integral for $\Jr$ in terms of complete elliptic integrals \citep[see][]{BF1971}:
\begin{equation}
\label{ap:JrE}
\begin{array}{l}
 \ds J_r = \frac{\sqrt{2GMa}}{\pi} D_5 \left[ D_1 \Pi \left( \alpha_1, k^2 \right) +
	      D_2 {\rm E}\left(k^2\right) + \right. \\ \\
 \ds \quad \left. + D_3 {\rm K}\left(k^2\right) +
	      D_4 \Pi \left( \alpha_2, k^2 \right) \right],
\end{array}
\end{equation}
where ${\rm K},{\rm E},\Pi$ are respectively the complete elliptic integral of the first, second and third kind,
\begin{equation}
\label{ap:alpha_kappa}
\alpha_1\equiv\frac{\Psi_1-\Psi_2}{\Psi_2},\quad \alpha_2\equiv\frac{\Psi_1-\Psi_2}{1-\Psi_2}, 
\quad k^2\equiv\frac{\Psi_1-\Psi_2 }{A-\Psi_2}
\end{equation}
and finally
\begin{equation}
\label{eq:coeffs}
\begin{array}{l}
 \displaystyle D_1 = [(1-2\Psi_1)\Psi_2+\Psi_1]A+\Psi_1\Psi_2, \\
 \displaystyle D_2 = \Psi_2(\Psi_2-A), \\
 \displaystyle D_3 = \Psi_2(A-2), \\
 \displaystyle D_4 = 2\Psi_2(A-1)(\Psi_1-1), \\
 \displaystyle D_5 = -\sqrt{A-\Psi_2}/D_2.
\end{array}
\end{equation}
We have tested the formula \eqref{ap:JrE} for consistency by numerically integrating equation \eqref{ap:def_Jr}
for a large set of orbits at different $(E,L)$ and the numerical and analytical results agree within the
error of the employed routine.

\end{document}